\documentclass[12pt]{elsart}

\usepackage{graphicx}
\usepackage{subfigure}
\usepackage{amsmath}
\usepackage{cite}
\usepackage{tikz}
\usepackage{setspace}

\journal{J. Magn. Magn. Mater.}

\begin{document}

\begin{frontmatter}

\title{Magnetization process and magnetocaloric effect in geometrically frustrated Ising antiferromagnet and spin ice models on a `Star of David' nanocluster}
\author{M. \v{Z}ukovi\v{c}\corauthref{cor}},
\ead{milan.zukovic@upjs.sk}
\author{M. Semjan}
\address{Institute of Physics, Faculty of Science, P.J. \v{S}af\'arik University, Park Angelinum 9, 041 54 Ko\v{s}ice, Slovakia}
\corauth[cor]{Corresponding author.}

\begin{abstract}
Magnetic and magnetocaloric properties of geometrically frustrated antiferromagnetic Ising (IA) and ferromagnetic spin ice (SI) models on a nanocluster with a `Star of David' topology, including next-nearest-neighbor (NNN) interactions, are studied by an exact enumeration. In an external field applied in characteristic directions of the respective models, depending on the NNN interaction sign and magnitude, the ground state magnetization of the IA model is found to display up to three intermediate plateaus at fractional values of the saturation magnetization, while the SI model shows only one zero-magnetization plateau and only for the antiferromagnetic NNN coupling. A giant magnetocaloric effect is revealed in the the IA model with the NNN interaction either absent or equal to the nearest-neighbor coupling. The latter is characterized by abrupt isothermal entropy changes at low temperatures and infinitely fast adiabatic temperature variations for specific entropy values in the processes when the magnetic field either vanishes or tends to the critical values related to the magnetization jumps.  
\end{abstract}


\end{frontmatter}

\section{Introduction}
\begin{figure}[t]
\centering
\includegraphics[scale=0.4,clip]{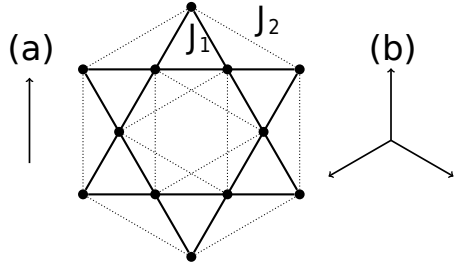}
\caption{`Star of David' nanocluster with NN (solid lines $J_1$) and NNN (dashed lines $J_2$) couplings. Local anisotropy axes in the direction of (a) $z$-axis for IA and (b) elementary triangle axes for SI models.}\label{fig:star_of_David}
\end{figure}

Molecular nanomagnets, composed of a finite number of interacting spins (spin cluster) magnetically decoupled from their environment, in recent decades have attracted considerable attention~\cite{kahn93,atta06,furr13}. They can be found in many real materials but nowadays they can also be artificially designed in a highly controlled manner~\cite{shen02,loun07,pede14}. In spite of their simplicity, such zero-dimensional magnetic structures are still of theoretical interest as they can provide an excellent opportunity to examine fundamental magnetic interactions, in case of small clusters, as well as possibilities to explore novel many-body quantum states and quantum phenomena, in case of larger clusters. From the practical point of view, they can find applications in data storage, quantum computing and molecule-based spintronics devices~\cite{leue01,boga08}. It has also been demonstrated that molecular nanomagnets can potentially show a very large magnetocaloric effect (MCE) at low temperatures (for an overview, see, e.g.,~\cite{evan06}) and are thus attractive materials for a magnetic cooling technology.\\
\hspace*{5mm} Even larger MCE can be achieved by the presence of degenerate or low-lying spin states that can be induced, for example, by frustration. It has been theoretically concluded that the field-dependent efficiency of a geometrically frustrated magnet can exceed that of an ideal paramagnet with equivalent spin by more than an order of magnitude~\cite{zhit1,schn07} and experimentally supported by the results obtained for the geometrically frustrated ${\rm Fe}_{14}$~\cite{evan05,shaw07} and ${\rm Ga}_{7}$~\cite{sharp14} molecular nanomagnets. Besides the enhanced MCE, an effect of frustration has been demonstrated to lead to a variety of other unusual magnetic properties, such as non-collinear ground states, magnetization plateaus and magnetization jumps~\cite{schn06,schn10}. Some of these properties have been studied on small antiferromagnetic Ising spin clusters~\cite{viit97} or more complex nanoparticles~\cite{zaim09,wei12,wei13,wei14,kant13}, focusing on the magnetization process. More recently investigations have been extended also to geometrically frustrated antiferromagnetic Ising spin clusters of different shapes and sizes on a triangular lattice~\cite{milla1,zuko14,zuko15} as well as to the clusters of the shapes of regular polyhedra (Platonic solids)~\cite{stre15,karl16}, focusing on both the magnetization and the adiabatic demagnetization processes. In both frustrated systems it was shown that the magnetization process (number and height of the magnetization plateaus) as well as the adiabatic demagnetization process (magnetocaloric properties) strongly depend on the cluster geometry. Several shapes displaying enhanced (giant) MCE were identified. We note that modern techniques enable syntheses of molecular magnets of various structures, thus providing opportunity for the selected systems showing giant MCE to be practically used in technological applications.\\
\hspace*{5mm} In the present study we focus on the magnetic and magnetocaloric properties of two geometrically frustrated spin systems: antiferromagnetic Ising (IA) and ferromagnetic spin ice (SI) models, including the effect of both the nearest-neighbor (NN) and the next-nearest-neighbor (NNN) interactions, localized on a nanocluster of a `Star of David' shape (Fig.~\ref{fig:star_of_David}). The latter is the elementary cell of a kagom\'{e} lattice, which, based on the  residual entropy as a measure of frustration, is considered to be the most frustrated lattice with no long-range ordering at any temperature~\cite{will02}. We note that very recently a family of single-molecule magnets with the `Star of David' topology has been synthesized and their magnetocaloric properties have been investigated~\cite{alex16}. Our study shows that both IA and SI systems display favorable magnetocaloric properties. Nevertheless, particularly fast cooling rates to ultra low temperatures at small fields were observed in the former system, which makes it a good candidate for application as an efficient low-temperature refrigerator. 

\section{Models} 

\subsection{Ising antiferromagnet}
The NNN IA model on the `Star of David' nanocluster in an external magnetic field can be described by the Hamiltonian 
\begin{equation}
\label{Hamiltonian1}
\mathcal H_{IA}=-J_1\sum_{\langle i,j \rangle}\sigma_{i}\sigma_{j} -J_2\sum_{\langle i,k \rangle}\sigma_{i}\sigma_{k} - h\sum_{i}\sigma_{i},
\end{equation}
where $\sigma_{i}=\pm 1/2$ are the Ising variables on the $i$th site, the summations $\langle i,j \rangle$ and $\langle i,k \rangle$ run over NN and NNN spin pairs, $J_1<0$ is an antiferromagnetic NN exchange interaction parameter, the NNN exchange interaction parameter is restricted to the range $-|J_1|<J_2<|J_1|$, and $h$ is the external magnetic field. The `Star of David' cluster consists of 6 corner-sharing triangles with $N=12$ Ising spins, arranged as shown in Fig.~\ref{fig:star_of_David}. Each of the spins can be in spin-up or spin-down states totaling in $2^{12}$ possible configurations.
\subsection{Spin ice}
Also in the SI model the spins possess the Ising anisotropy, however, the Ising axes now run through the center of the triangular plaquettes (Fig.~\ref{fig:star_of_David}) and the interaction between the neighboring spins is ferromagnetic. The Hamiltonian of the corresponding NNN SI model reads 
\begin{equation}
\label{Hamiltonian2a}
\mathcal H_{SI}=-J_1\sum_{\langle i,j \rangle}{\bf s}_{i}{\bf s}_{j} -J_2\sum_{\langle i,k \rangle}{\bf s}_{i}{\bf s}_{k} - h\sum_{i}{\bf e}_h{\bf s}_{i},
\end{equation}
where $J_1>0$, $-J_1<J_2<J_1$, and ${\bf e}_h$ is a unit vector along the field direction. The spin ice variables can be related to the Ising variables by the relation ${\bf s}_{i}=\sigma_i{\bf e}_i$, with ${\bf e}_i$ being a unit vector along the easy axis, which for the SI model is the line connecting the centers of triangles.

\section{Exact enumeration}
Taking advantage of the possibility to explore the entire state space of the system consisting of a relatively small number of spins, one can exactly determine the density of states and consequently calculate all the thermodynamic quantities of interest. \\
\hspace*{5mm} Ground states (GS) in either system are found for different field values as the configurations that minimize the energy functionals~(\ref{Hamiltonian1}) and~(\ref{Hamiltonian2a}). Then, we count the number of GS configurations $W$ (degeneracy) and evaluate the corresponding magnetizations $M_j=\sum_{i=1}^{N}{\sigma_{i}{\bf e}_i{\bf e}_h}$, $j=1,\ldots,W$, where ${\bf e}_i{\bf e}_h=1$ for the IA model, as functions of the external field. Finally, we calculate the reduced magnetization $m=\sum_{j=1}^{W}{M_j}/(WN)$ and the entropy density $s=\ln W/N$ (here and hereafter, we put the Boltzmann constant $k_B=1$).\\
\hspace*{5mm} Having obtained the exact density of states $g(E_1,E_2,M)$, as a function of the total exchange energies $E_1=-J_1\sum_{\langle i,j \rangle}\sigma_{i}\sigma_{j}{\bf e}_i{\bf e}_j$ and $E_2=-J_2\sum_{\langle i,k \rangle}\sigma_{i}\sigma_{k}{\bf e}_i{\bf e}_k$, where the summations $\langle i,j \rangle$ and $\langle i,k \rangle$ run over NN and NNN on the considered cluster, respectively, and the total magnetization $M=\sum_{i=1}^{N}{\sigma_{i}{\bf e}_i{\bf e}_h}$, at finite temperatures one can calculate a mean value of an arbitrary thermodynamic quantity $A$, as a function the parameters $T$ and $h$, using the relation
\begin{equation}
\label{mean}
A(T,h)=\frac{\sum_{E_1,E_2,M}{Ag(E_1,E_2,M)e^{-\frac{(E_1+E_2-hM)}{T}}}}{Z(T,h)},
\end{equation}
where 
\begin{equation}
\label{PF}
Z(T,h)=\sum_{E_1,E_2,M}{g(E_1,E_2,M)e^{-\frac{(E_1+E_2-hM)}{T}}} 
\end{equation}
is the partition function. Then, the magnetization per spin can be obtained as 
\begin{equation}
\label{magnetization}
m(T,h)=\langle M \rangle/N
\end{equation}
and the entropy density as
\begin{equation}
\label{entropy}
s(T,h)=\frac{E(T,h)-F(T,h)}{NT},
\end{equation}
where $E(T,h)=\langle E_1+E_2-hM \rangle$ is the enthalpy and $F(T,h)=-T\ln{Z(T,h)}$ is the free energy. Without loss of generality, in the following we can set $|J_1|=1$ and then consider the reduced variables $J_2/|J_1|=J_2$, $k_BT/|J_1|=T$ and $h/|J_1|=h$.

\section{Results}
\subsection{Ising antiferromagnet}

\subsubsection{Ground state}

\begin{figure}[t]
\centering
\subfigure{\includegraphics[scale=0.35,clip]{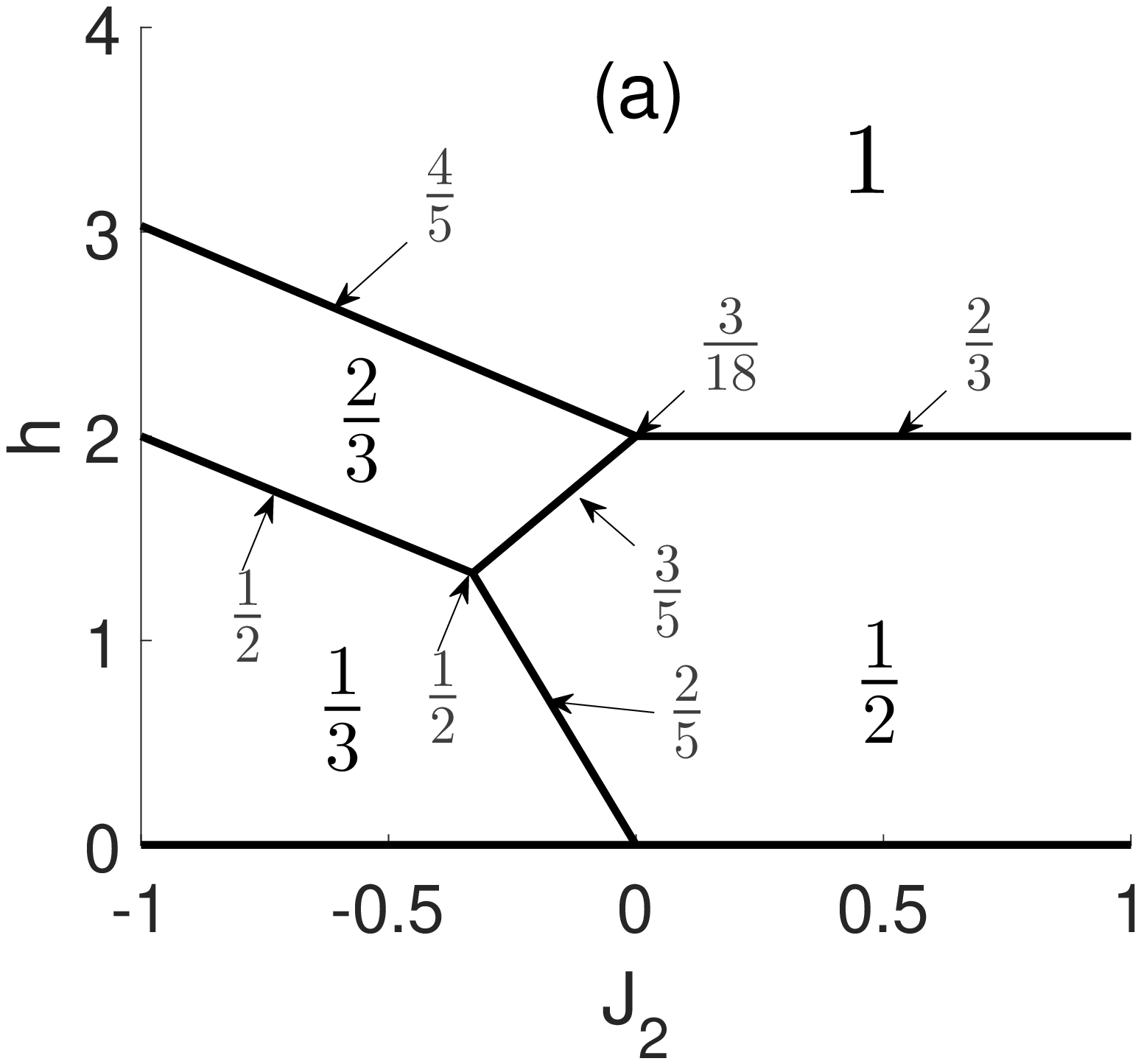}\label{fig:m_GS_IA}}
\subfigure{\includegraphics[scale=0.35,clip]{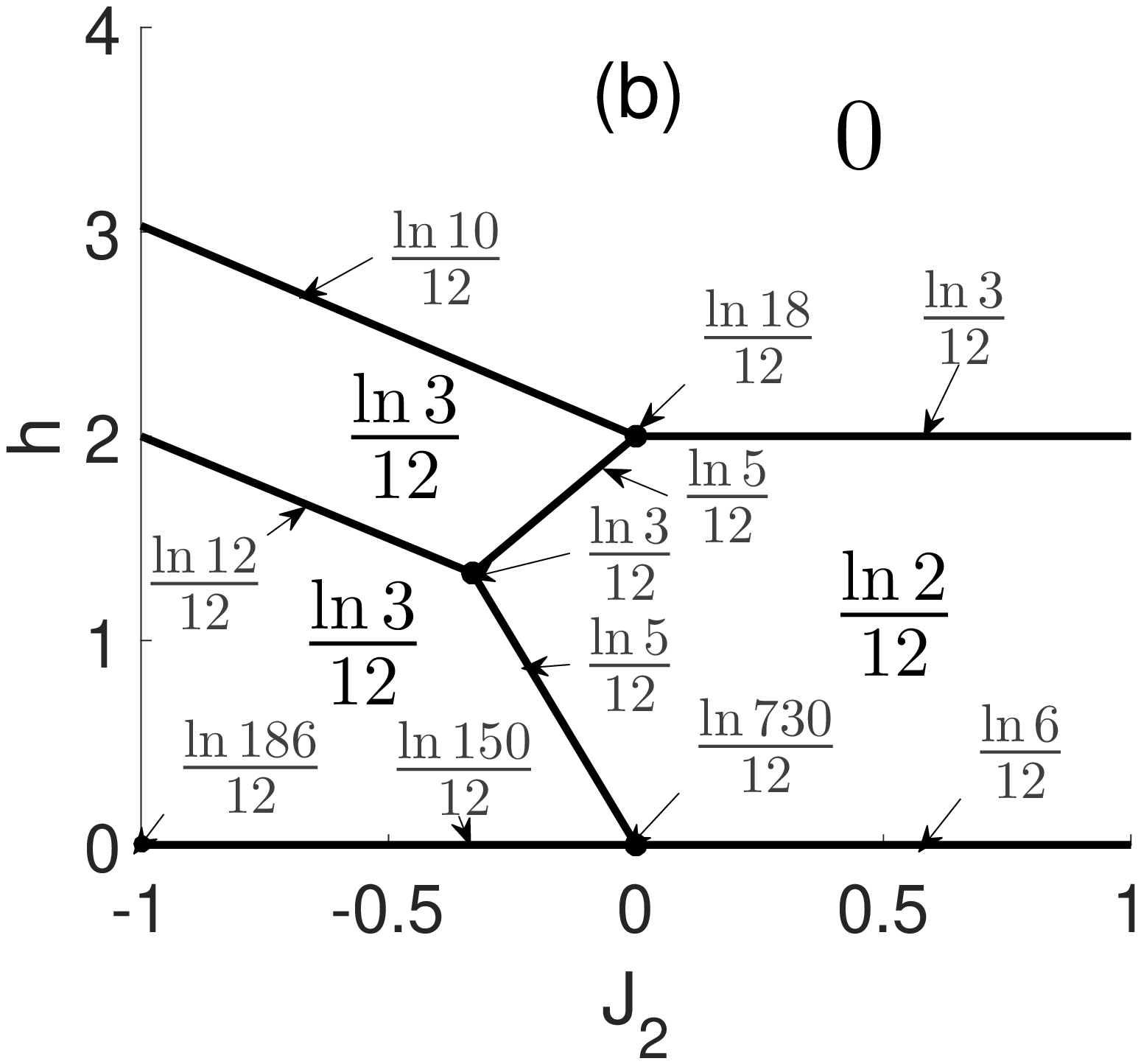}\label{fig:s_GS_IA}}
\subfigure{\includegraphics[scale=0.35,clip]{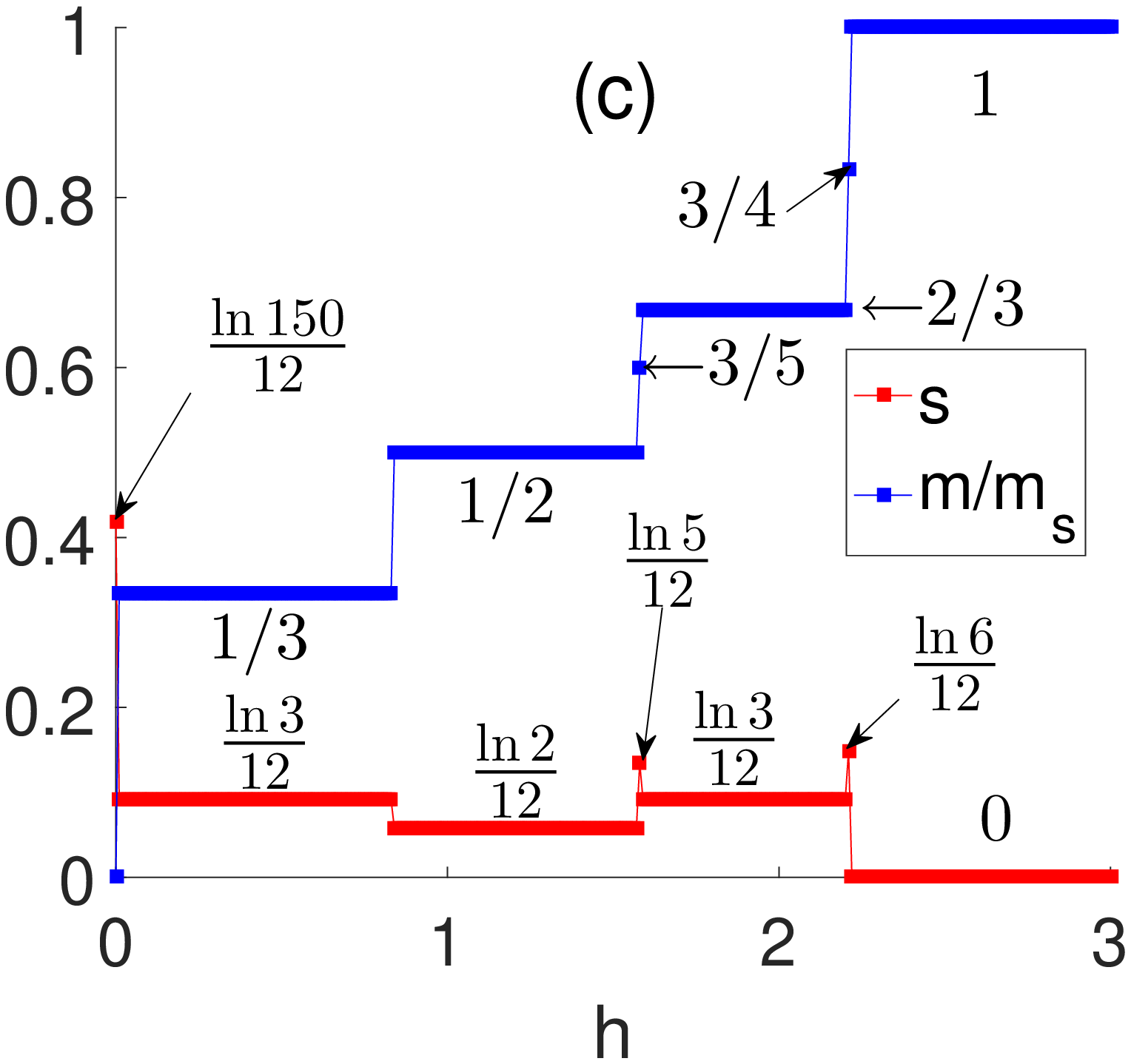}\label{fig:m_s_GS_IA}}
\caption{Ground state (a) normalized magnetization $m/m_s$ and (b) entropy density $s$ of the IA model in the $(h,J_2)$ parameter plane. Values shown in larger font (smaller font with arrows) correspond to the interior (boundaries) of the respective phases. (c) Magnetization and entropy processes, for $J_2=-0.2$.}\label{fig:GS_IA}
\end{figure}

The GS values of the magnetization, normalized with respect to the saturation value $m_{s}=1/2$, and the entropy density $s$ are depicted in Fig.~\ref{fig:GS_IA} in the parameter plane $(h,J_2)=[0,h_{max}]\times[-1,1]$. In zero field, the magnetization (Fig.~\ref{fig:m_GS_IA}) is always vanishing. Depending on $J_2$, between the zero-field value $m/m_{s}=0$ and the saturation-field value $m/m_{s}=1$ the magnetization process can include up to three plateaus. Namely, for $J_2>0$ there is only one plateau of the height $1/2$ within $0<h<2$; for $J_2<-1/3$ there are two plateaus of the heights $1/3$ within $0<h<-J_2+1$ and $2/3$ within $-J_2+1<h<-J_2+2$; and for $-1/3<J_2<0$ there are three plateaus of the heights $1/3$ within $0<h<-4J_2$, $1/2$ within $-4J_2<h<2J_2+2$ and $2/3$ within $2J_2+2<h<-J_2+2$. The magnetization jumps occur at critical fields at which the Zeeman contribution in the Hamiltonian (\ref{Hamiltonian1}), corresponding to the portion of the spins with the weakest coupling, overcomes their exchange energy. \\
\hspace*{5mm} The GS degeneracies $W$ in the $(h,J_2)$ parameter plane are apparent from the entropy density values presented in Fig.~\ref{fig:s_GS_IA} in the form $\ln W/N$. In zero field, $W$ depends on $J_2$ as follows: $W(J_2=0)=730$, $W(J_2=-1)=186$, $W(-1<J_2<0)=150$, and $W(J_2>0)=6$. In a finite field, the degeneracies pertaining the the magnetization plateaus $m/m_{s}=1/3,1/2$ and $2/3$ are $W=3,2$ and $3$, respectively. It is interesting to notice that, in spite of the intuitive expectation that the increasing field should gradually decrease $W$, within $-1/3<J_2<0$ above the critical field $h_c=2J_2+2$ the degeneracy actually increases before being completely lifted at the saturation field $h_s=-J_2+2$. \\
\hspace*{5mm} Note that both the magnetization and entropy values at the respective critical fields differ from the values inside the plateaus, as shown in Fig~\ref{fig:GS_IA}. The entropy in locally increased due to contributions of degenerate states coming from two or even three neighboring plateaus and the values of $m/m_{s}$ correspond to degeneracy-weighted averages of the neighboring plateaus values. For illustration, the magnetization and entropy processes for $J_2=-0.2$, that include three intermediate plateaus, are presented in Fig.~\ref{fig:m_s_GS_IA}. 

\subsubsection{Finite temperatures}

\begin{figure}[t]
\centering
\subfigure{\includegraphics[scale=0.23,clip]{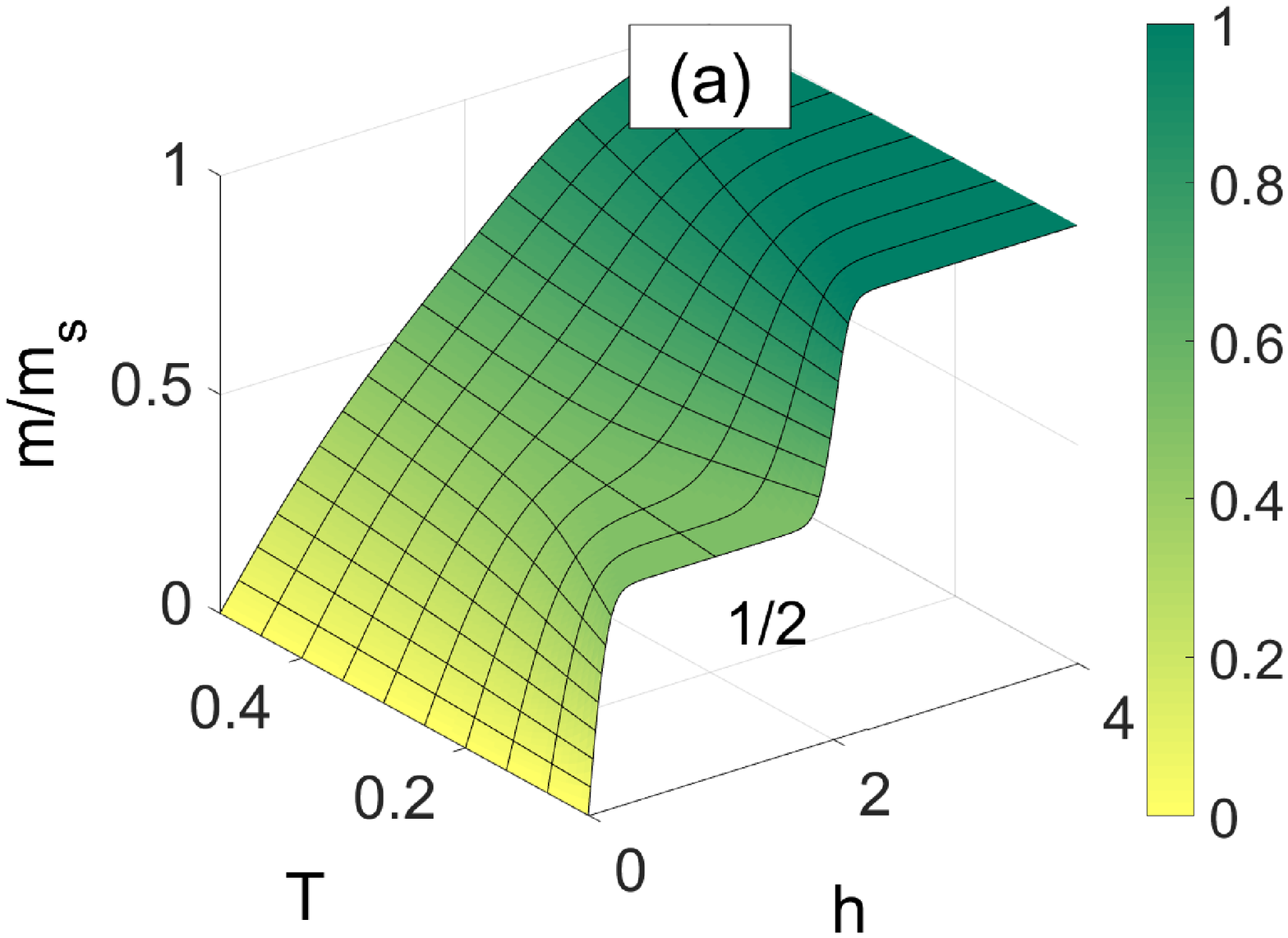}\label{fig:m_FT_IA_J0}}
\subfigure{\includegraphics[scale=0.23,clip]{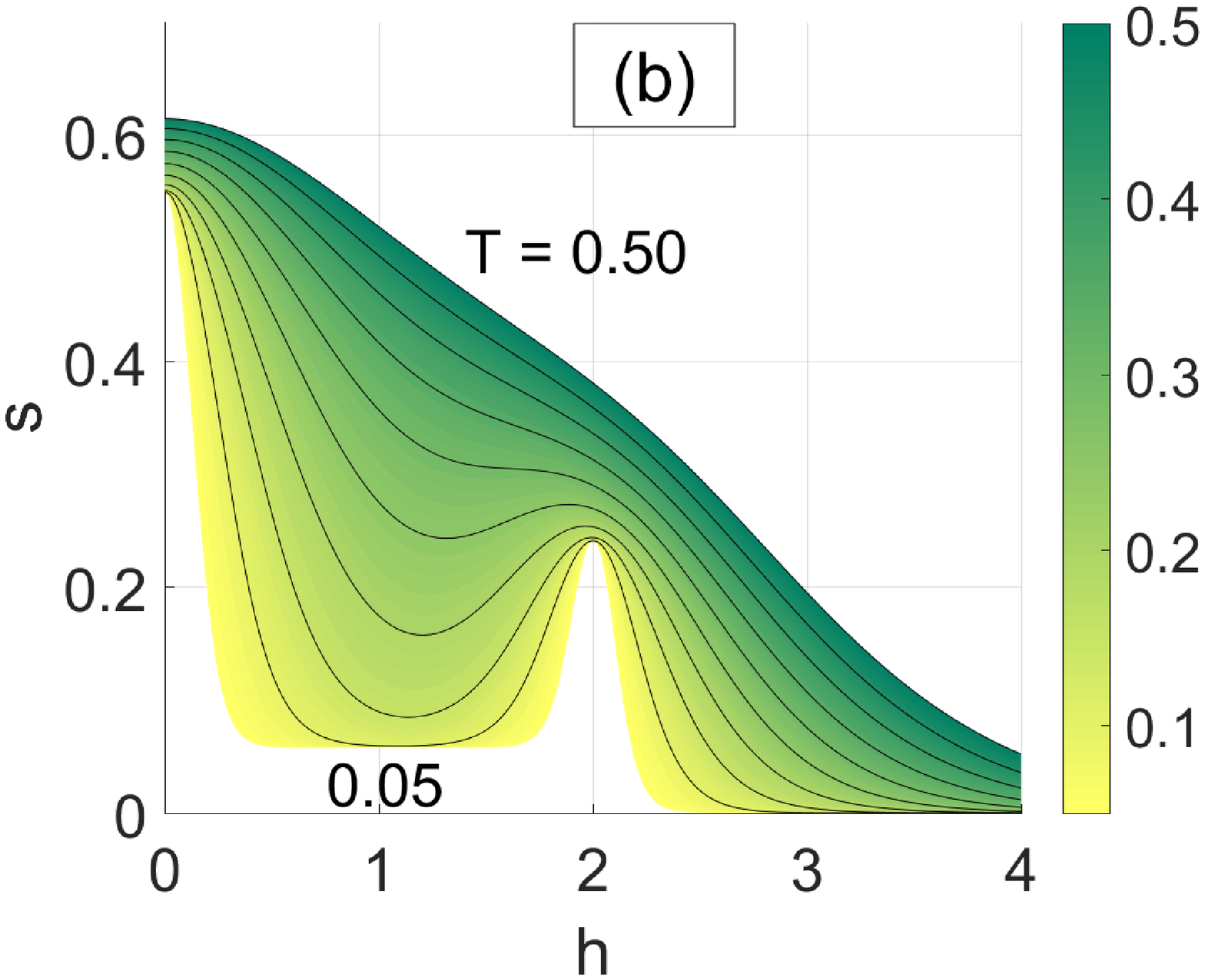}\label{fig:ds_FT_IA_J0}}
\subfigure{\includegraphics[scale=0.23,clip]{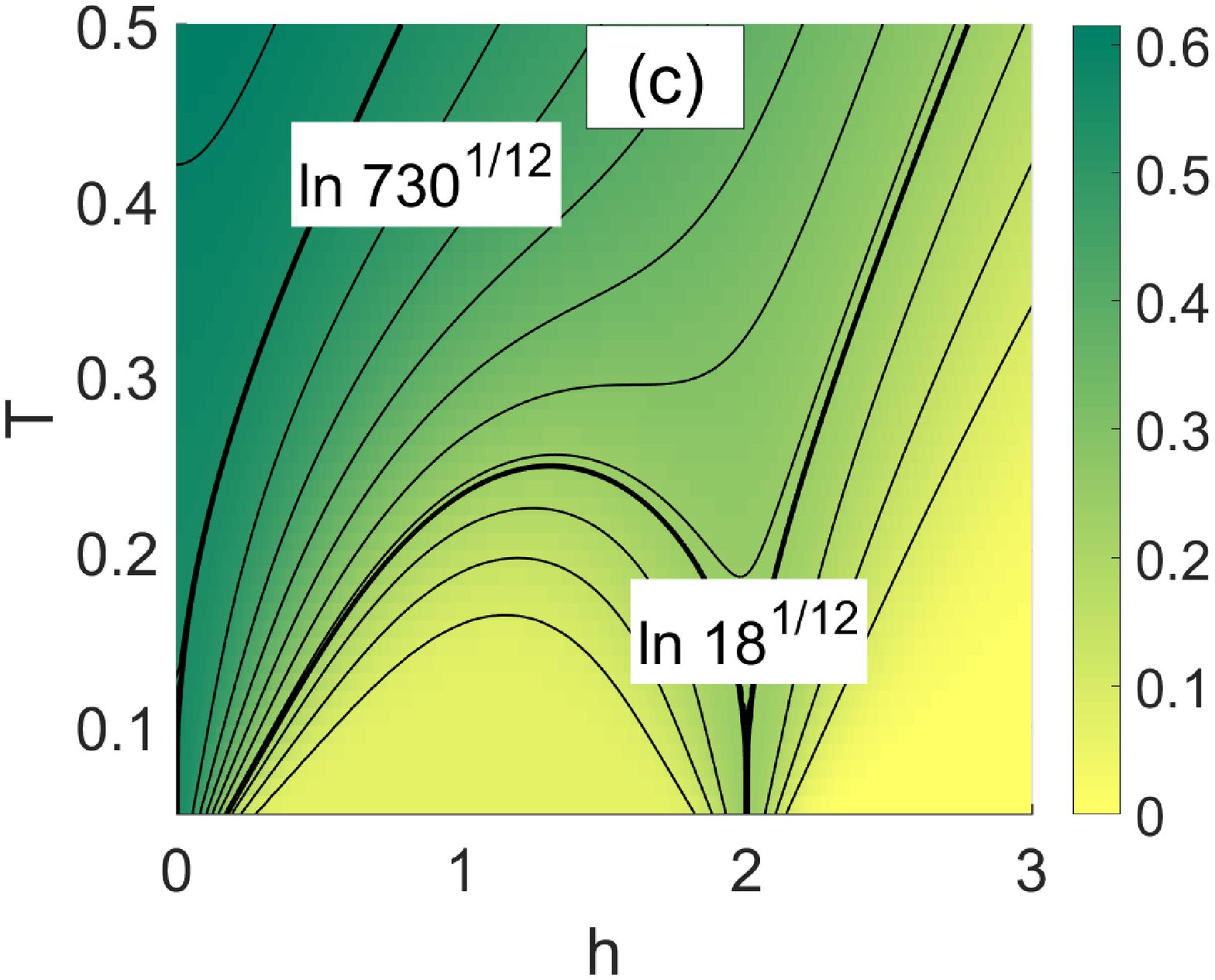}\label{fig:dt_FT_IA_J0}}\\
\subfigure{\includegraphics[scale=0.23,clip]{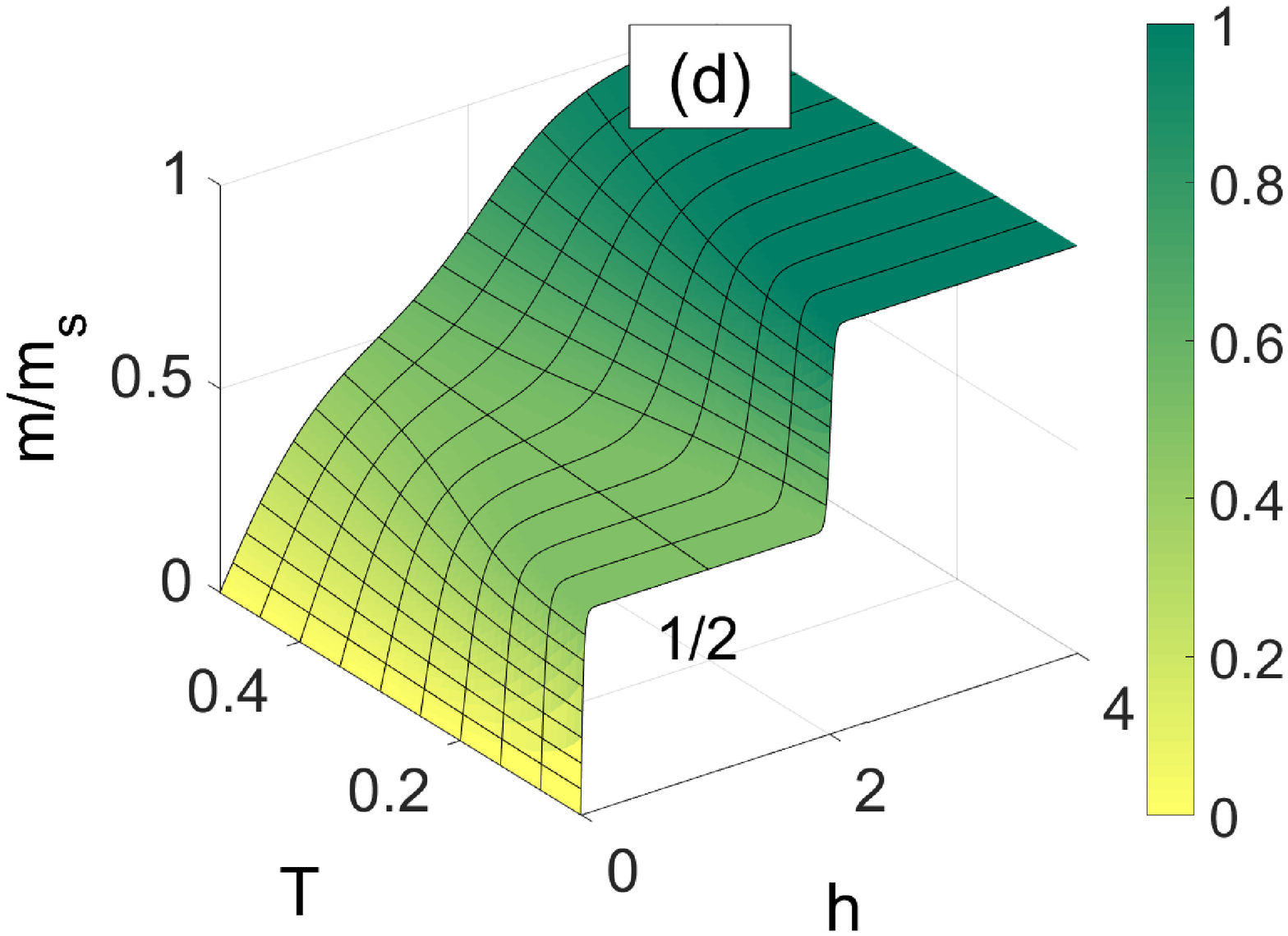}\label{fig:m_FT_IA_J05}}
\subfigure{\includegraphics[scale=0.23,clip]{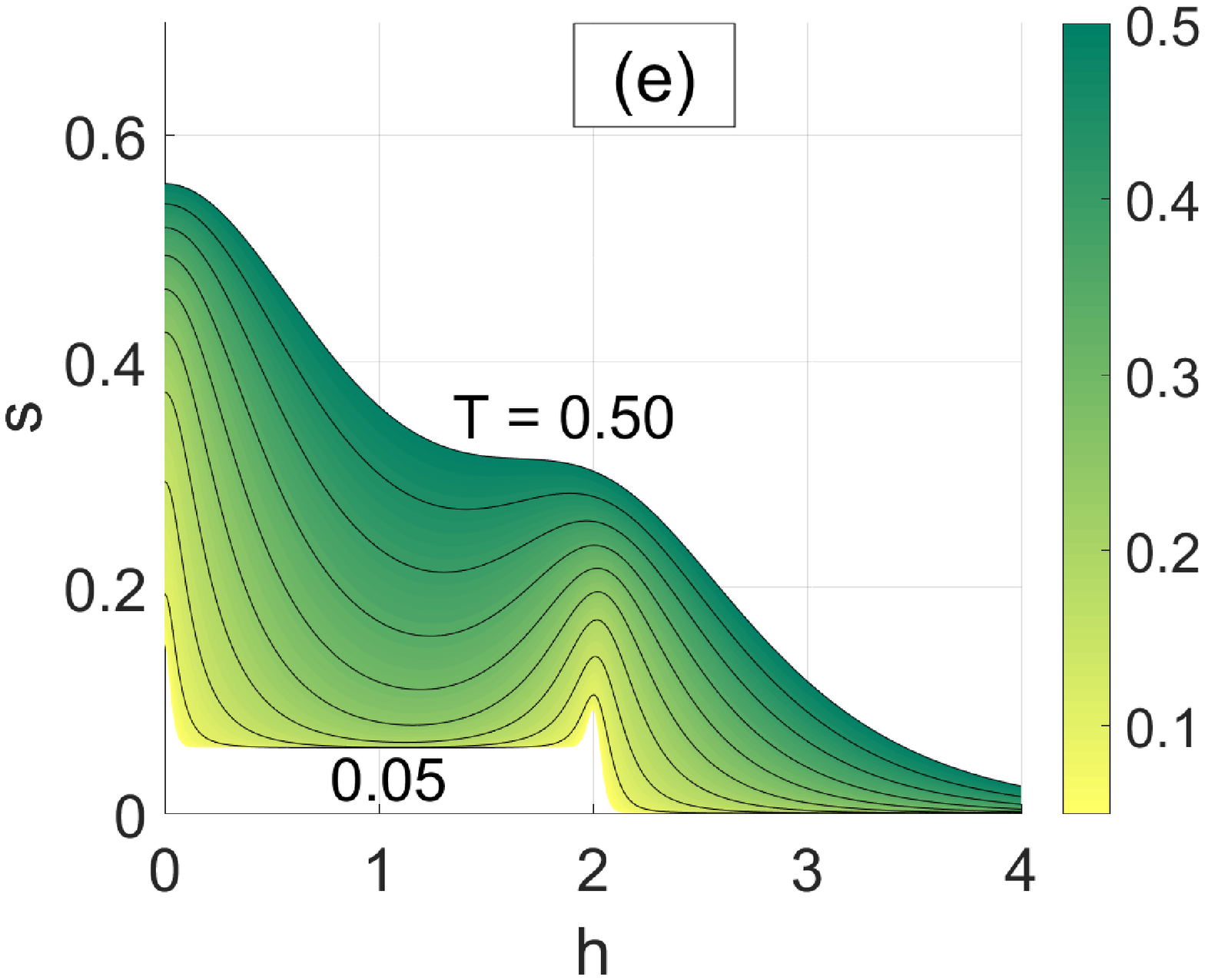}\label{fig:ds_FT_IA_J05}}
\subfigure{\includegraphics[scale=0.23,clip]{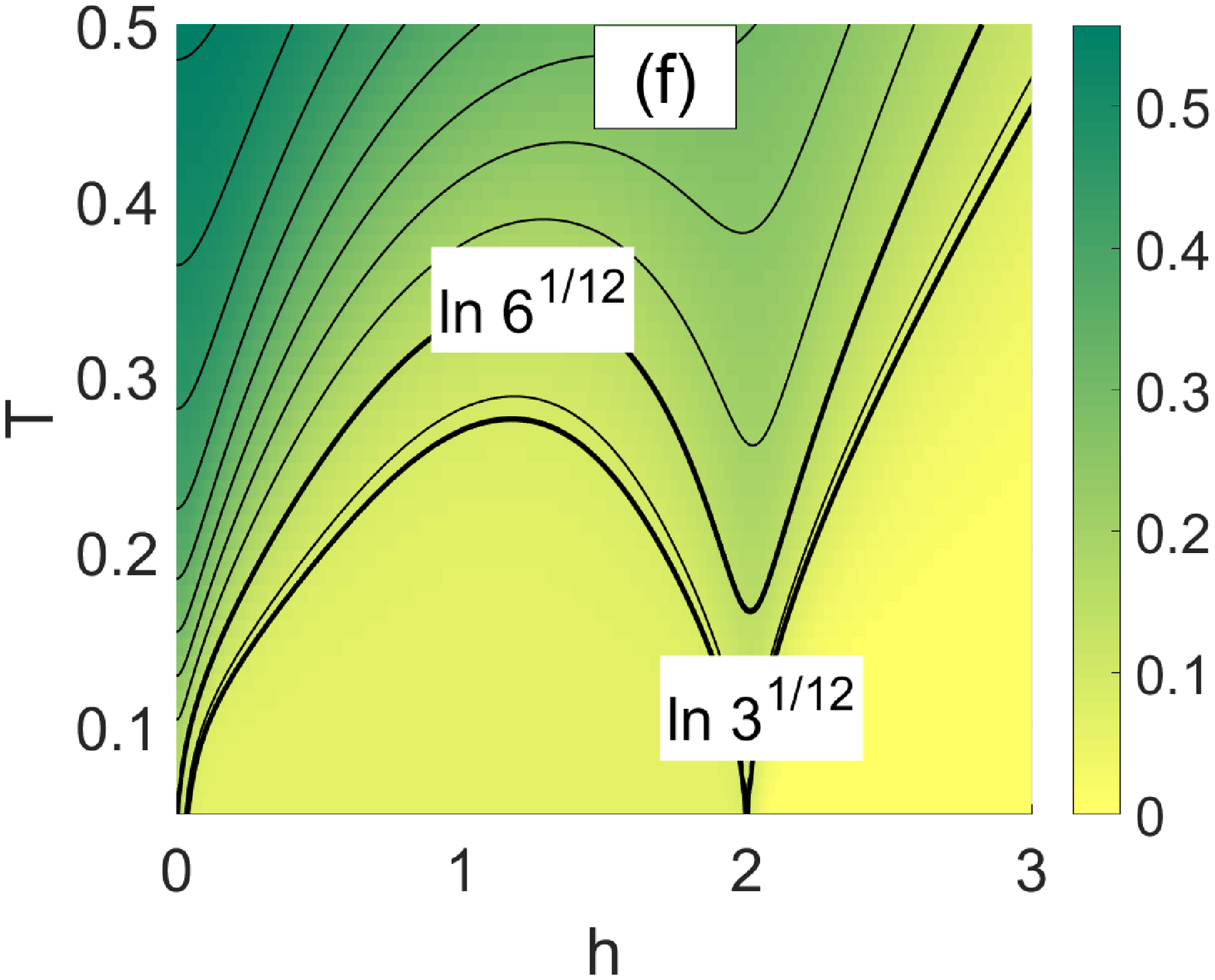}\label{fig:dt_FT_IA_J05}}\\
\subfigure{\includegraphics[scale=0.23,clip]{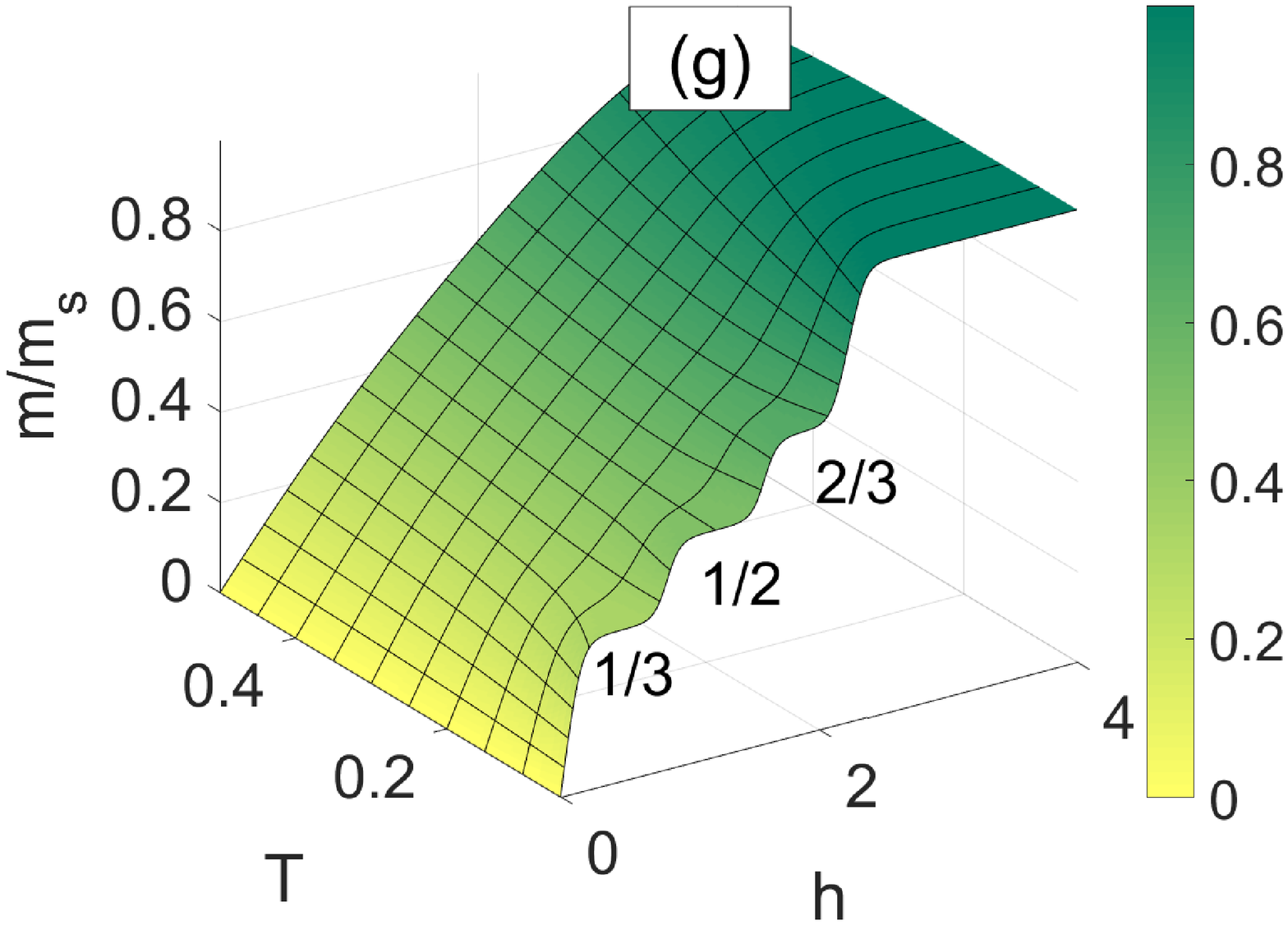}\label{fig:m_FT_IA_J-02}}
\subfigure{\includegraphics[scale=0.23,clip]{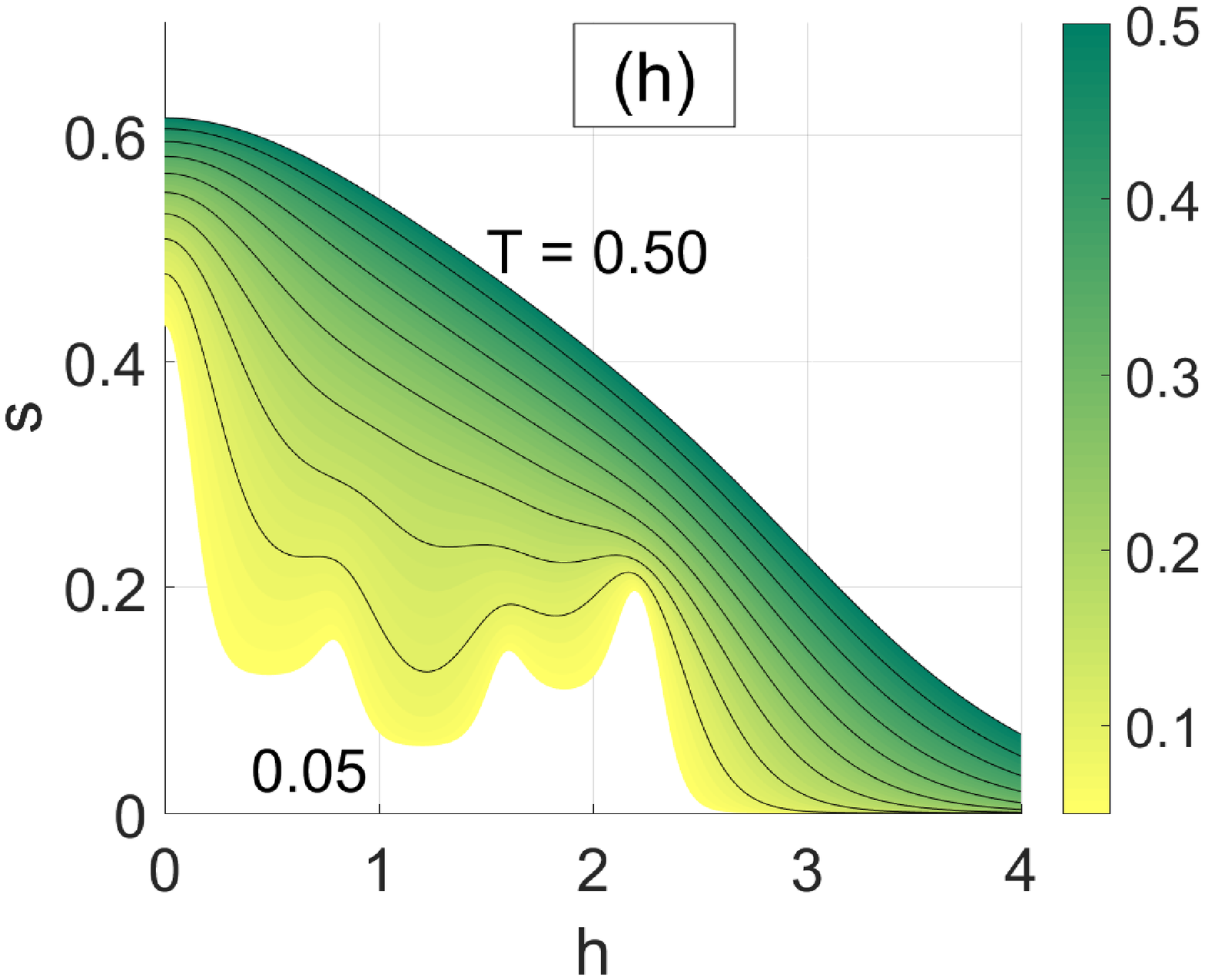}\label{fig:ds_FT_IA_J-02}}
\subfigure{\includegraphics[scale=0.23,clip]{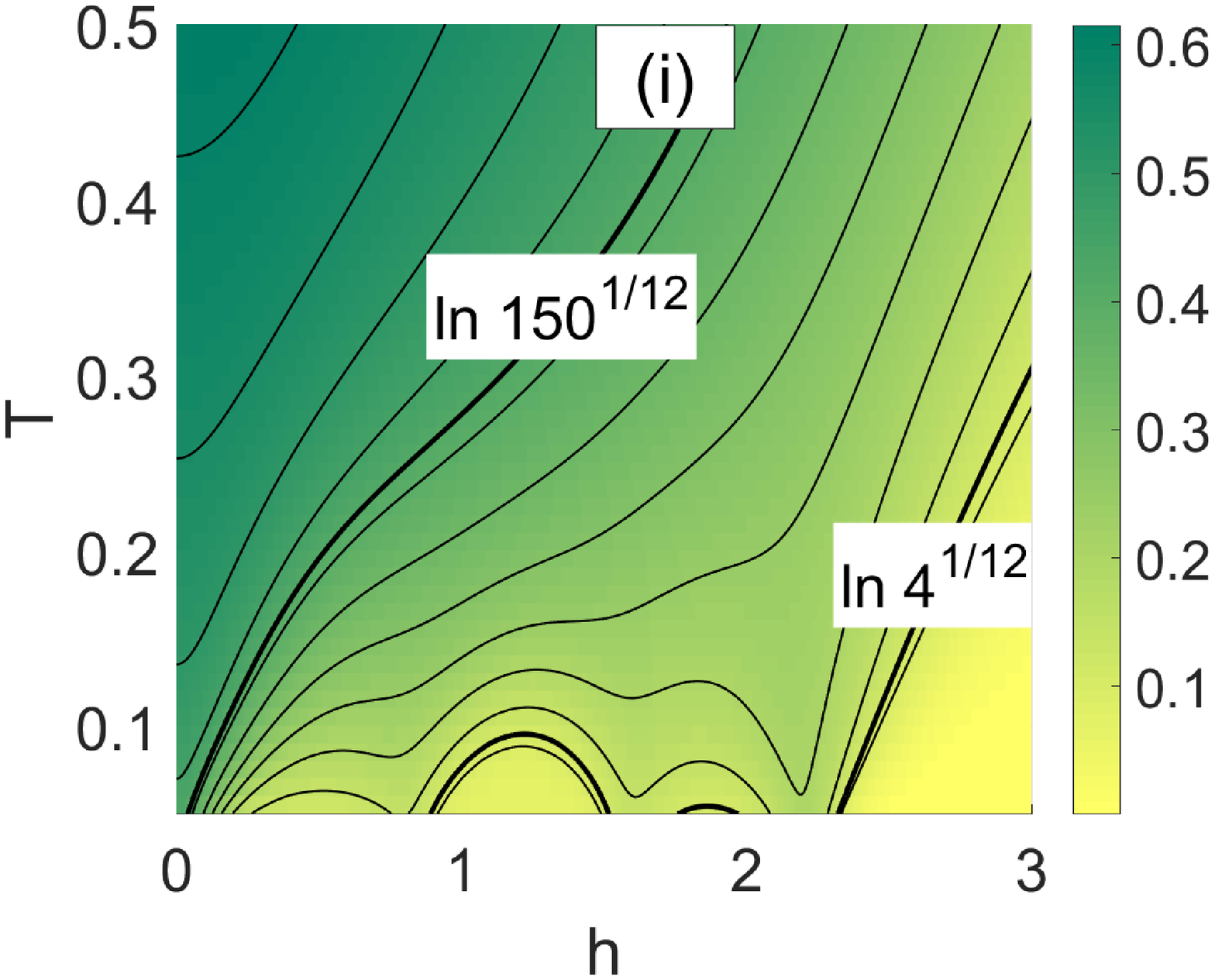}\label{fig:dt_FT_IA_J-02}}\\
\subfigure{\includegraphics[scale=0.23,clip]{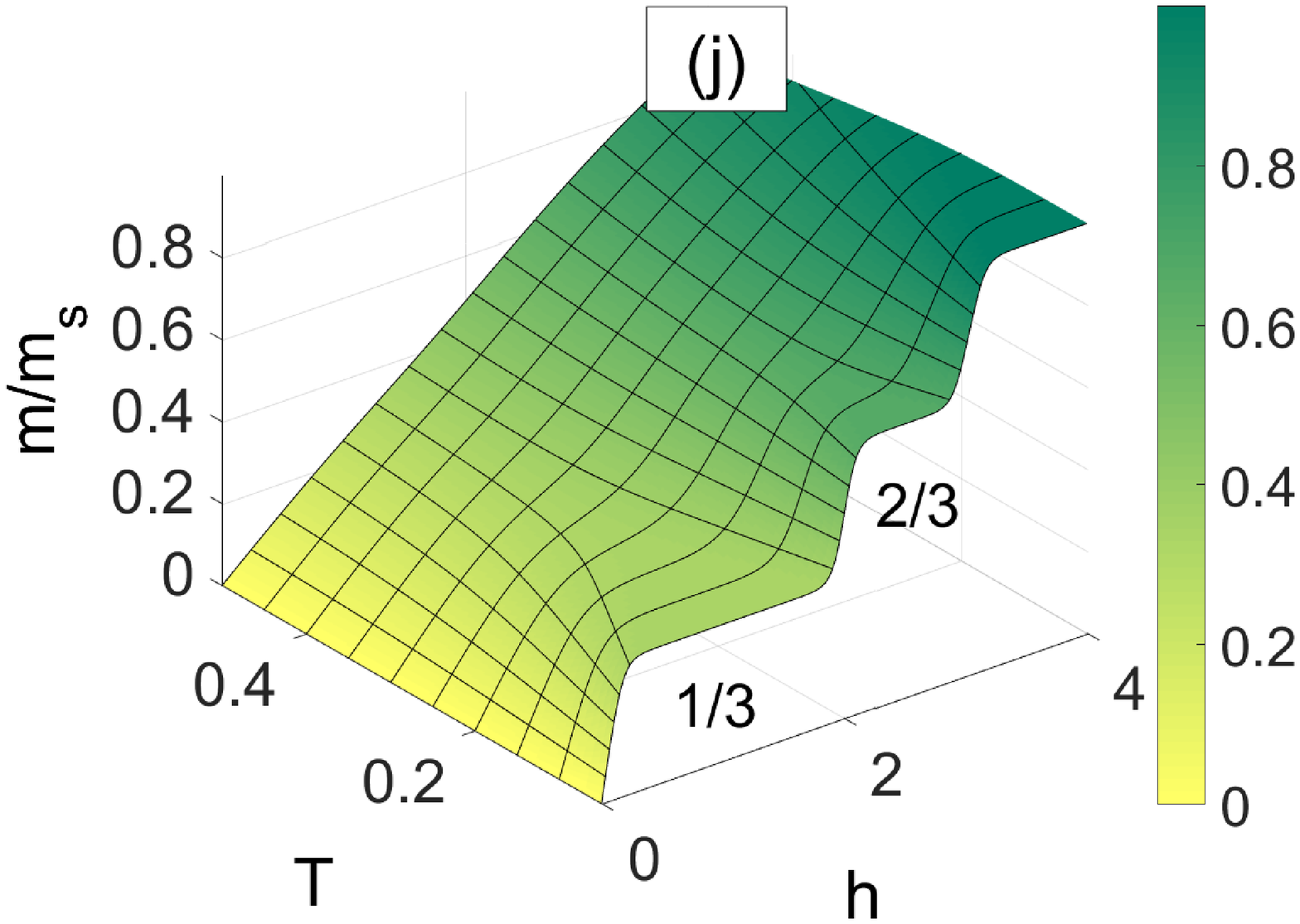}\label{fig:m_FT_IA_J-1}}
\subfigure{\includegraphics[scale=0.23,clip]{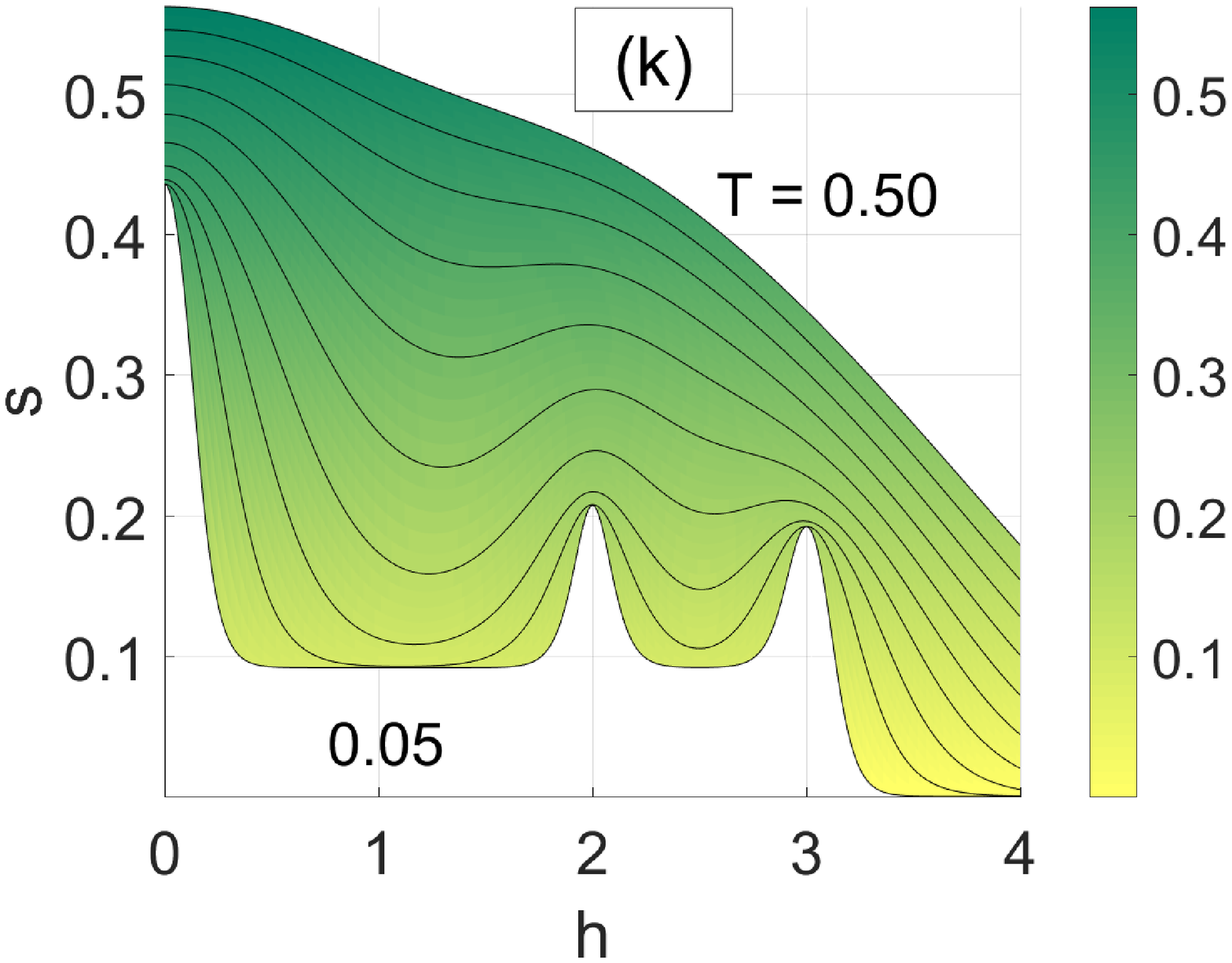}\label{fig:ds_FT_IA_J-1}}
\subfigure{\includegraphics[scale=0.23,clip]{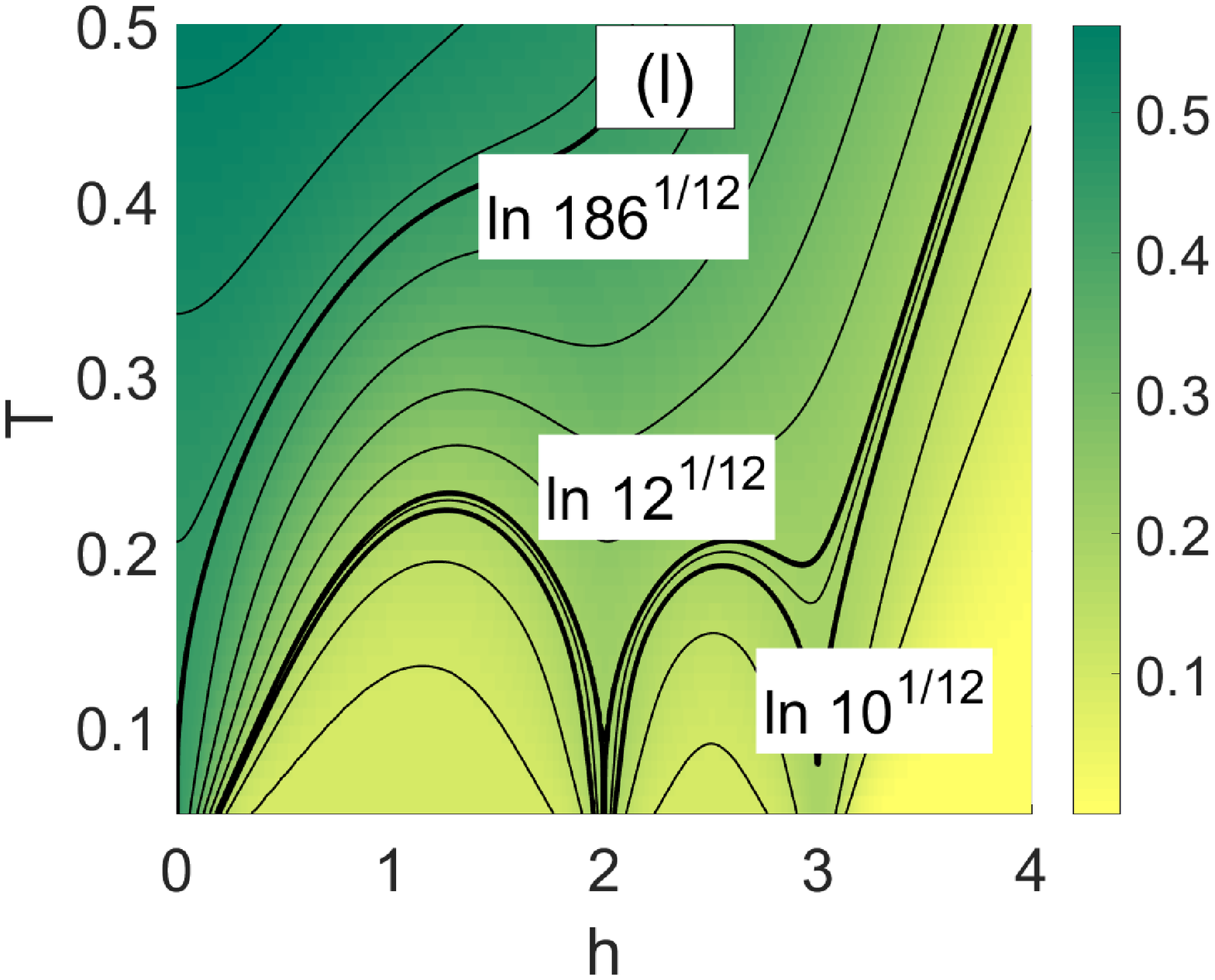}\label{fig:dt_FT_IA_J-1}}
\caption{Ising antiferromagnet: 3D plots of (a) the reduced magnetization $m/m_{s}$ and (b,c) the entropy density $s$ in the $(T,h)$ parameter plane. In (b) and (c) the black curves represent, respectively, isothermal entropy changes at different temperatures $T \in [0.05, 0.5]$, with the step $\Delta T=0.05$, and adiabatic temperature changes at various $s$ with a varying magnetic field $h$, for $J_2=0$. Panels (d-f), (g-i), and (j-l) show the same quantities as in (a-c), for $J_2=0.5$, $-0.2$, and $-1$, respectively. The bold curves in the right column represent isentropes corresponding to the GS entropies at the respective critical fields.}\label{fig:FT_IA}
\end{figure}

At finite temperatures, as expected, thermal fluctuations gradually smear out the sharp magnetizations steps observed at zero temperature, as demonstrated in Fig.~\ref{fig:FT_IA}, for $J_2=0, 0.5, -0.2$ and $-1$, respectively. More interesting are the magnetocaloric properties in the vicinity of the magnetization jumps, which are manifested by the isothermal entropy and adiabatic temperature changes. These can be respectively understood as the entropy and temperature responses to the magnetic field changes under isothermal and adiabatic conditions and, thus, can be represented by isotherms and insentropes on the entropy surface in the $(T,h)$ parameter plane. Fig.~\ref{fig:FT_IA} shows that dramatic changes of both quantities occur in the neighborhood of the critical fields, particularly for $J_2=0$ (Figs.~\ref{fig:ds_FT_IA_J0},~\ref{fig:dt_FT_IA_J0}), with the critical fields $h_1=0$ and $h_2=2$. In particular, the isotherms show drastic changes at low temperatures with very steep gradients for $T = 0$ as well as the isentropes close to the values corresponding to the residual entropies at the critical fields. The gradients of the latter tend to infinity for $s = \frac{\ln 730}{12}$, when $h$ approaches $h_1=0$ from above, and for $s = \frac{\ln 18}{12}$, when $h$ approaches $h_2=2$ from either side. Owing to the enhanced magnetocaloric properties observed in the vicinity of the magnetization jumps, the molecular nanomagnets that correspond to the present system can be potentially used in technological applications as magnetic refrigerants. For that purpose, particularly appealing is the extremely fast cooling to ultra low temperatures in the adiabatic demagnetization process with the vanishing magnetic field. Following the same line of considerations, one can arrive to the conclusion that generally the NNN interaction has an adverse effect on the magnetocaloric properties. As evidenced from Figs.~\ref{fig:ds_FT_IA_J05},~\ref{fig:dt_FT_IA_J05}, for $J_2=0.5$ and Figs.~\ref{fig:ds_FT_IA_J-02},~\ref{fig:dt_FT_IA_J-02}, for $J_2=-0.2$, the NNN coupling reduces the GS degeneracy and in the latter case also splits the intermediate magnetization plateau into three smaller steps and thus hampers the occurrence of changes in both quantities as dramatic as in the $J_2=0$ case. Nevertheless, the case of $J_2=-1$ is an exception showing the magnetocaloric properties similar to the case of $J_2=0$ (Figs.~\ref{fig:ds_FT_IA_J-1},~\ref{fig:dt_FT_IA_J-1}).

\subsection{Spin ice}

\subsubsection{Ground state}

\begin{figure}[t]
\centering
\subfigure{\includegraphics[scale=0.35,clip]{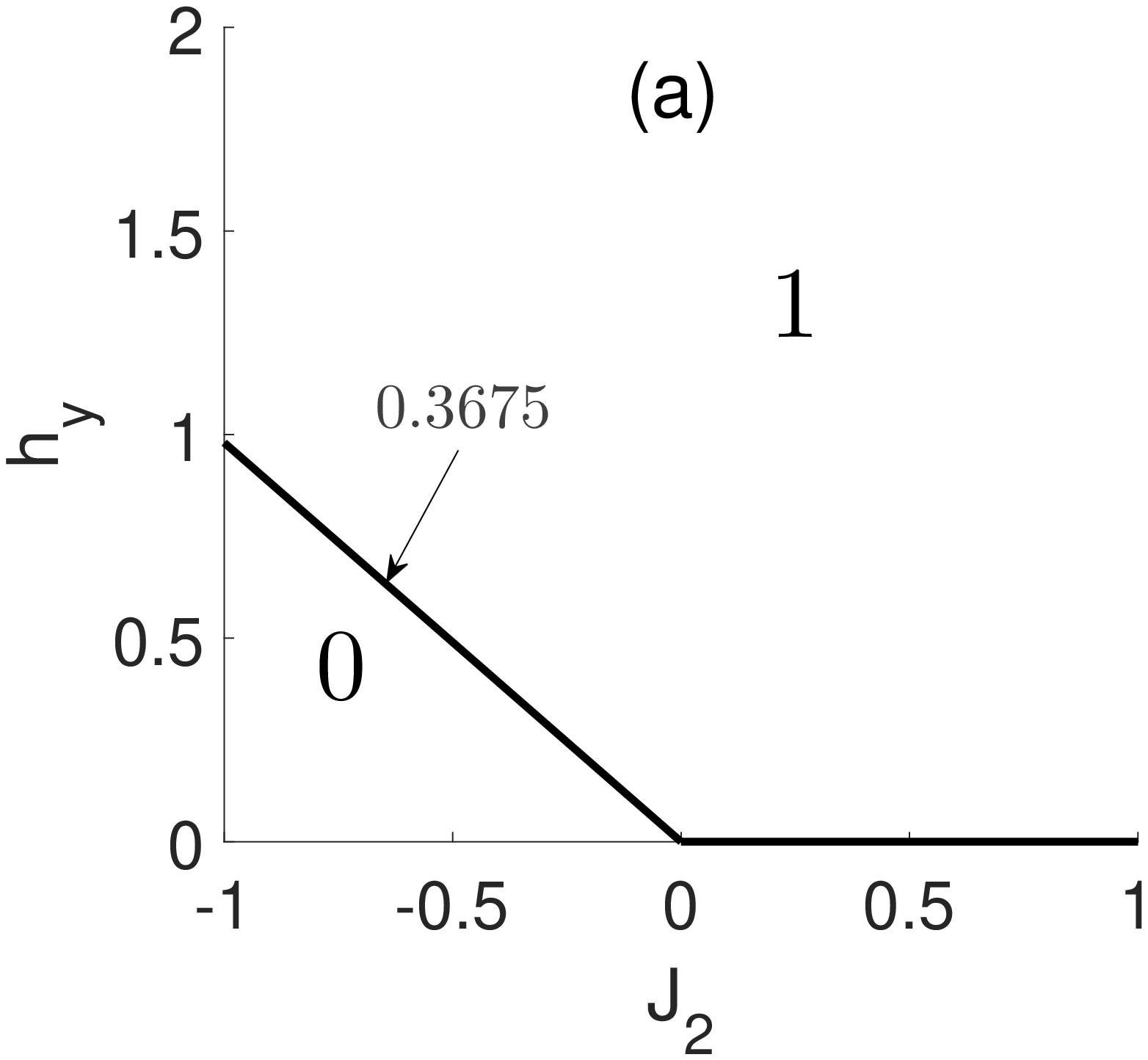}\label{fig:m_GS_SI_y}}
\subfigure{\includegraphics[scale=0.35,clip]{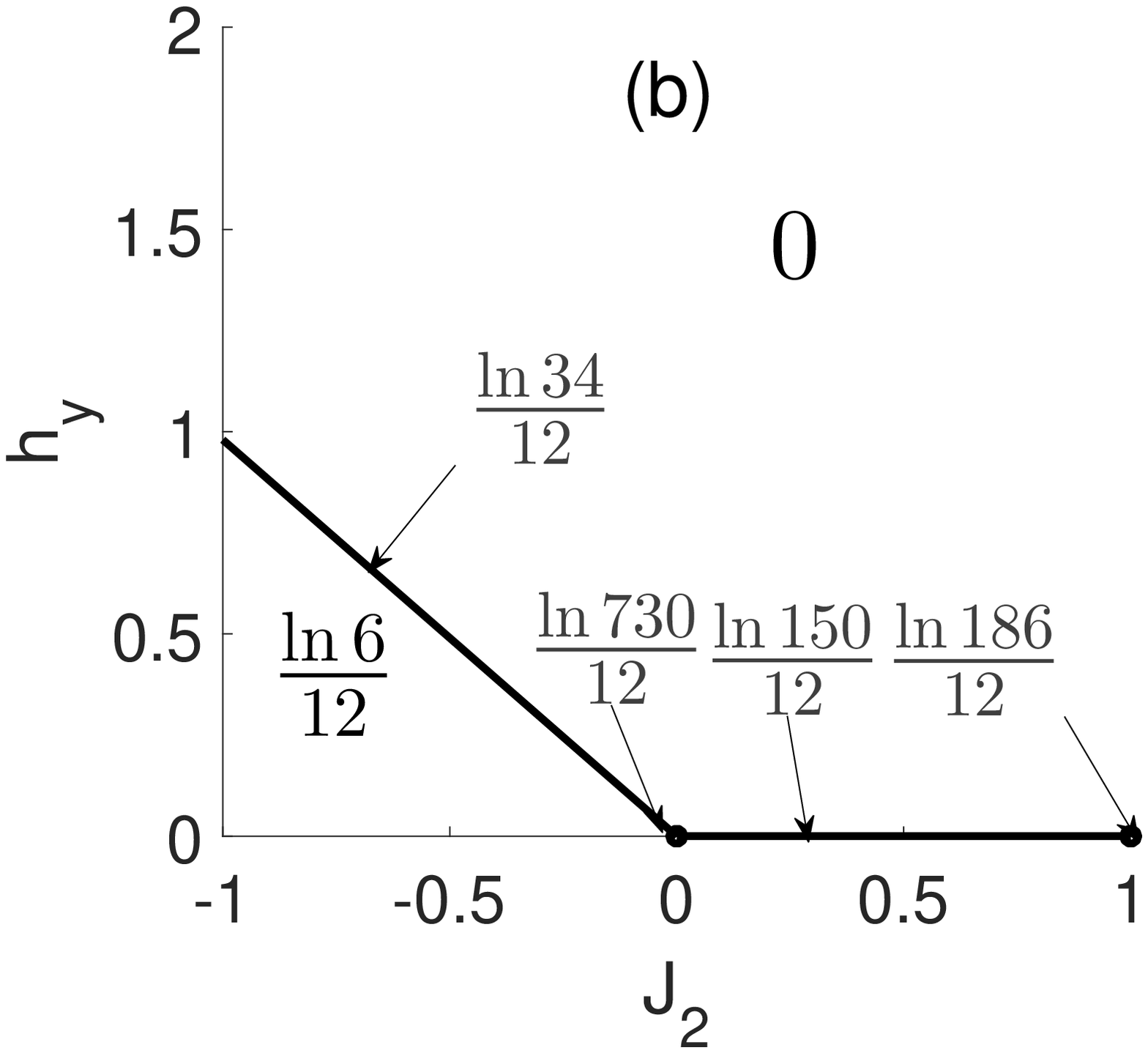}\label{fig:s_GS_SI_y}}
\subfigure{\includegraphics[scale=0.35,clip]{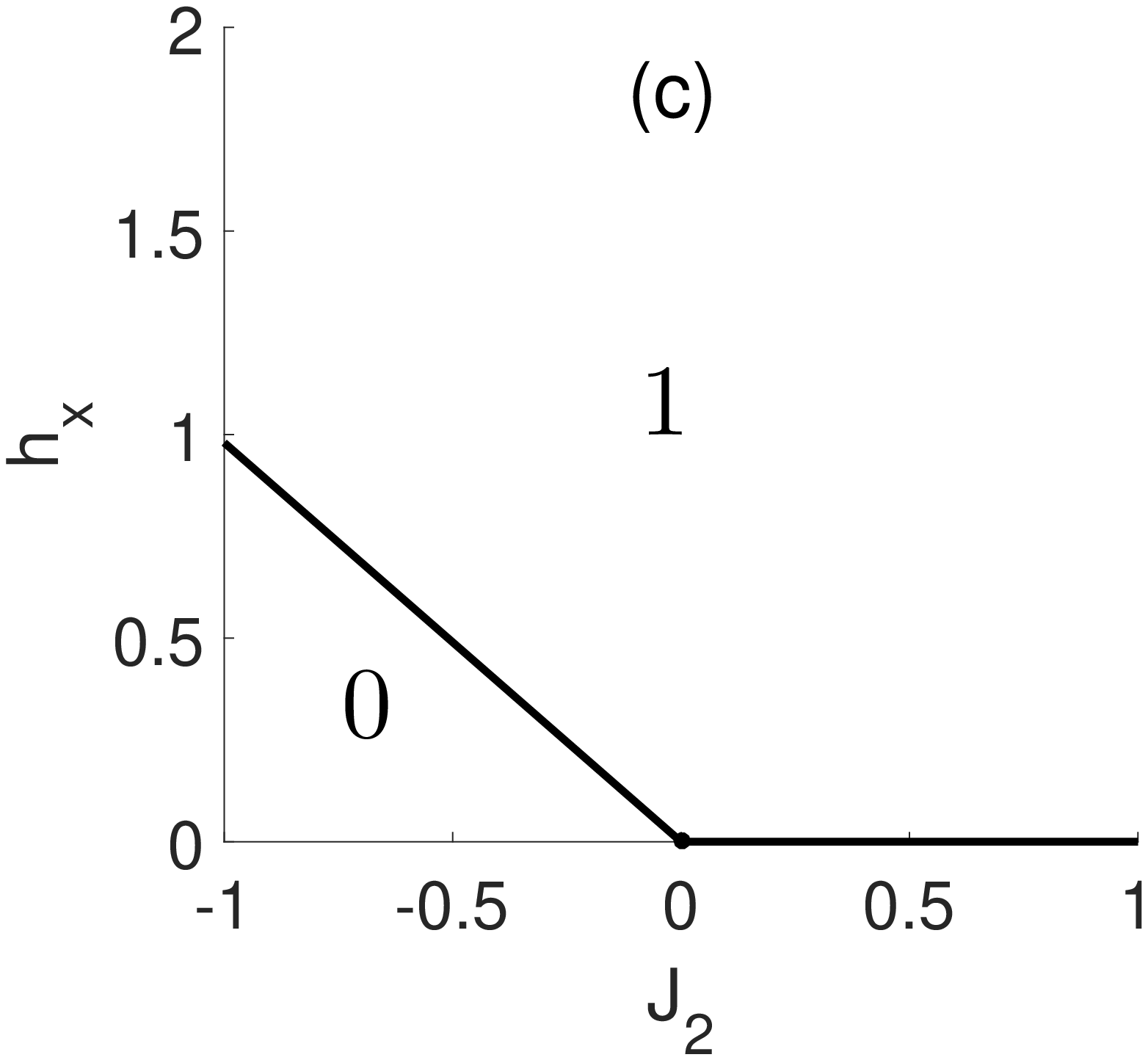}\label{fig:m_GS_SI_x}}
\subfigure{\includegraphics[scale=0.35,clip]{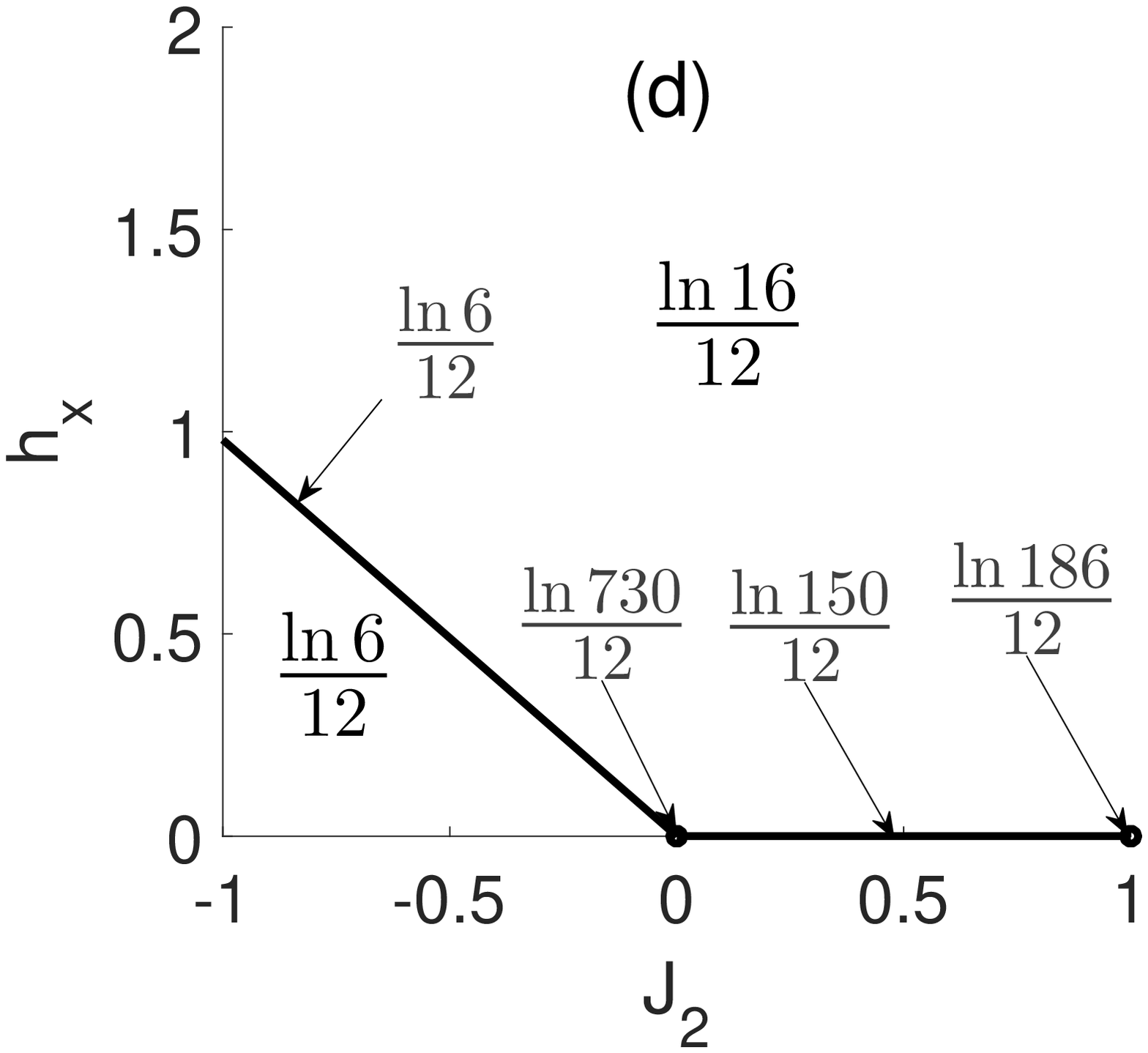}\label{fig:s_GS_SI_x}}
\caption{Ground state (a,c) normalized magnetization $m/m_s$ and (b,d) entropy density $s$ of the SI model in (a,b) $(h_y,J_2)$ and (c,d) $(h_x,J_2)$ parameter planes. Values shown in larger font (smaller font with arrows) correspond to the interior (boundaries) of the respective phases.}\label{fig:GS_SI}
\end{figure}

\begin{figure}[t]
\centering
\subfigure{\includegraphics[scale=0.3,clip]{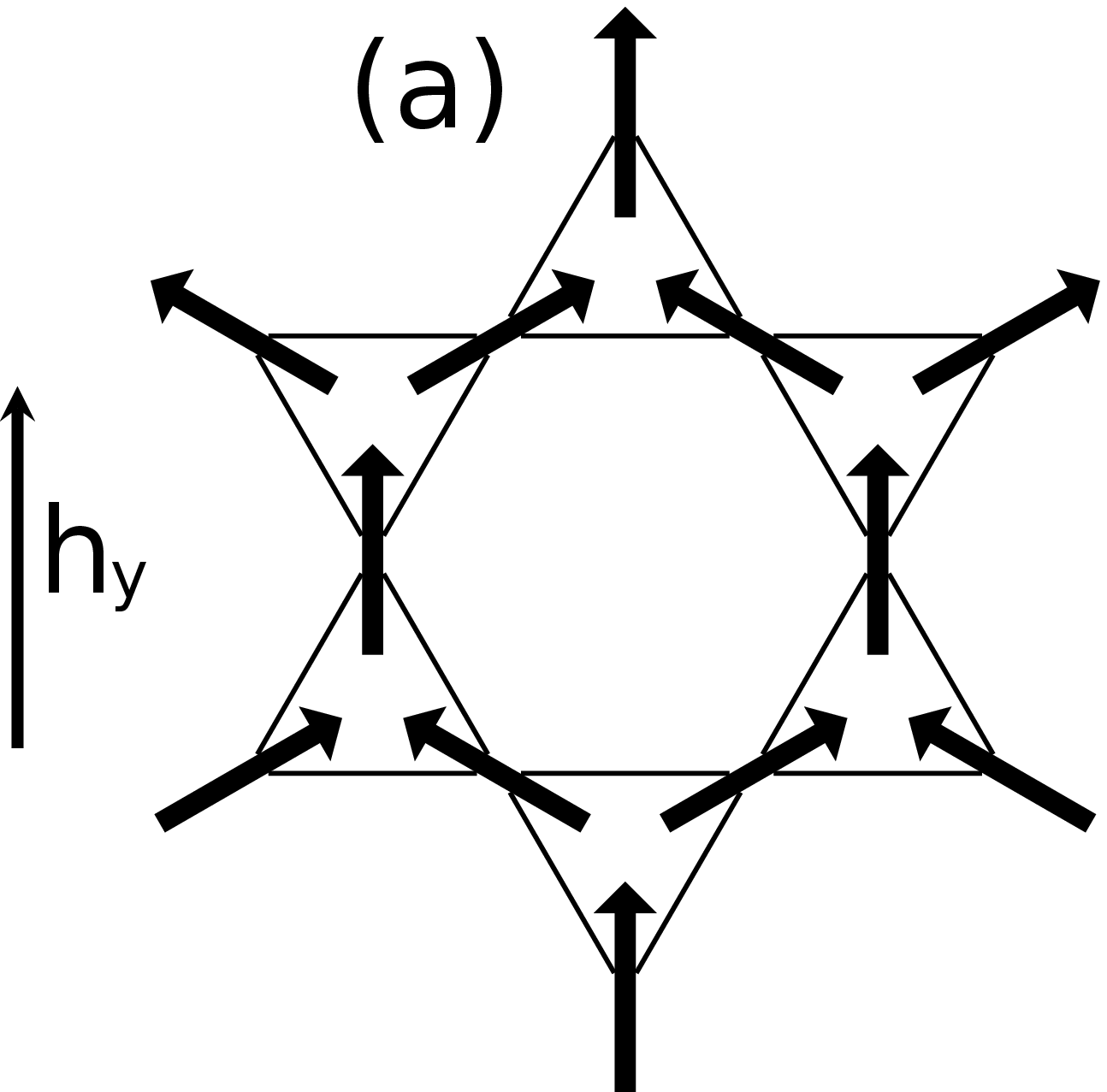}\label{fig:SI_vert}}\hspace*{10mm}
\subfigure{\includegraphics[scale=0.3,clip]{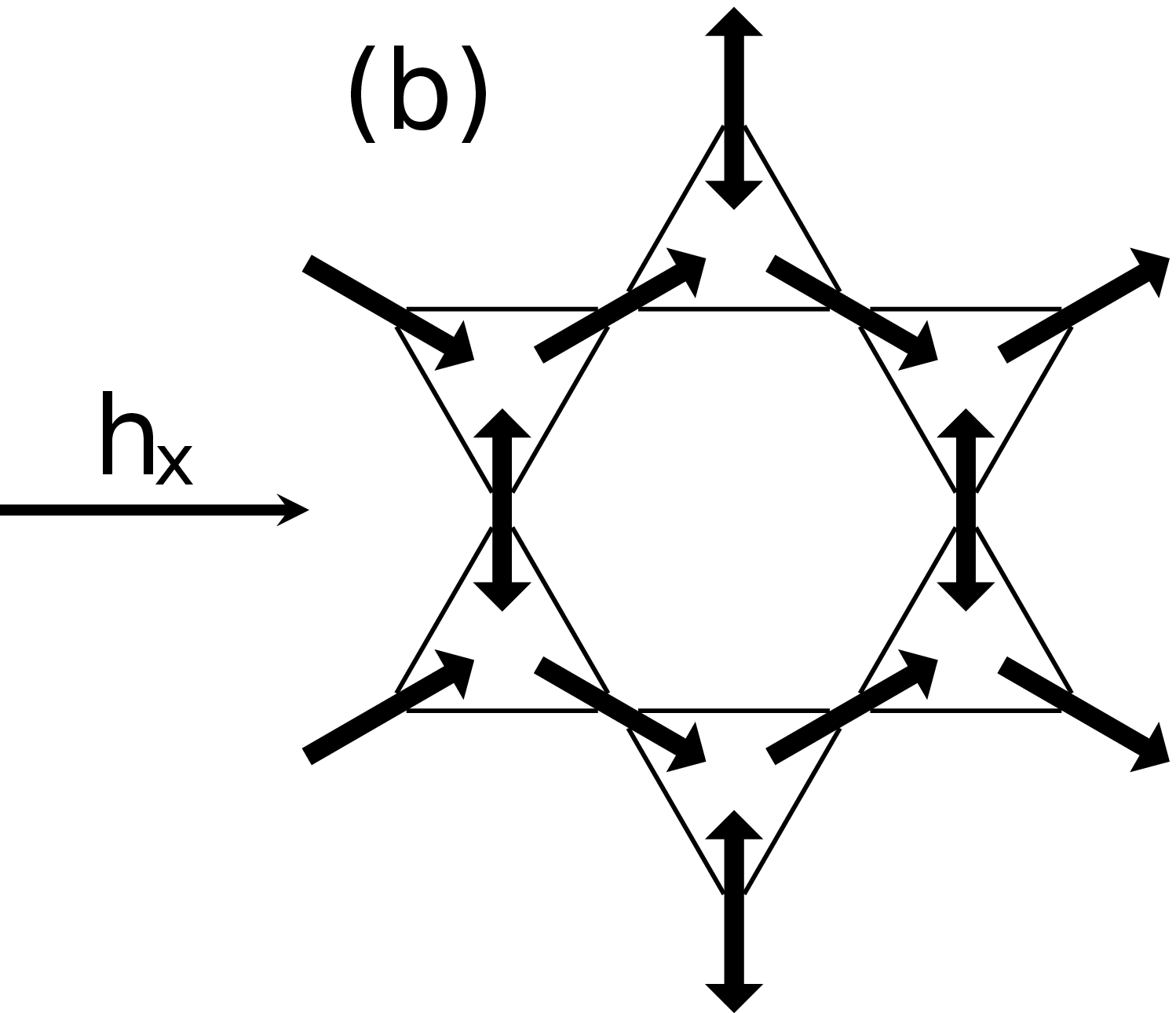}\label{fig:SI_hori}}
\caption{Saturation phase for the field applied (a) vertically and (b) horizontally. Double arrows represent the `free' spins.}\label{fig:free_spins}
\end{figure}

In zero field the spin ice Hamiltonian rewritten in terms of the Ising variables reduces to that for the IA model up the rescaled coupling constants differing by the factor of $-1/2$ and, therefore, no novel phenomena can be expected. However, in a finite field the behavior of the two models is not the same any more. In fact, it also depends on the applied field direction, even though the topology of the phase diagram does not change. This is evident from Fig.~\ref{fig:GS_SI}, for the fields $h_y$ and $h_x$ applied in the vertical (top row) and horizontal (bottom row) directions, respectively. In both cases, after application of the field the magnetizations either jump to the respective saturation values ($1/3$ for $h_y$ and $\sqrt{3}/6$ for $h_x$), if $J_2 \geq 0$, or show zero plateaus below the saturation field $h_s=-J_2$, if $J_2<0$. For $J_2<0$, the zero-field GS degeneracy $W=6$ remains the same in the zero magnetization plateau areas in both cases. However, in the case when the field is applied vertically it is completely lifted above the saturation field, while in the case when the field is applied horizontally it is even {\it increased} in the saturation phase. The resulting degeneracy in the latter case, $W=16=2^4$, corresponds to the number of possible states of the four `free' spins that can reverse their directions at no energy cost, as illustrated in Fig.~\ref{fig:free_spins}.

\subsubsection{Finite temperatures}
\begin{figure}[t]
\centering
\subfigure{\includegraphics[scale=0.23,clip]{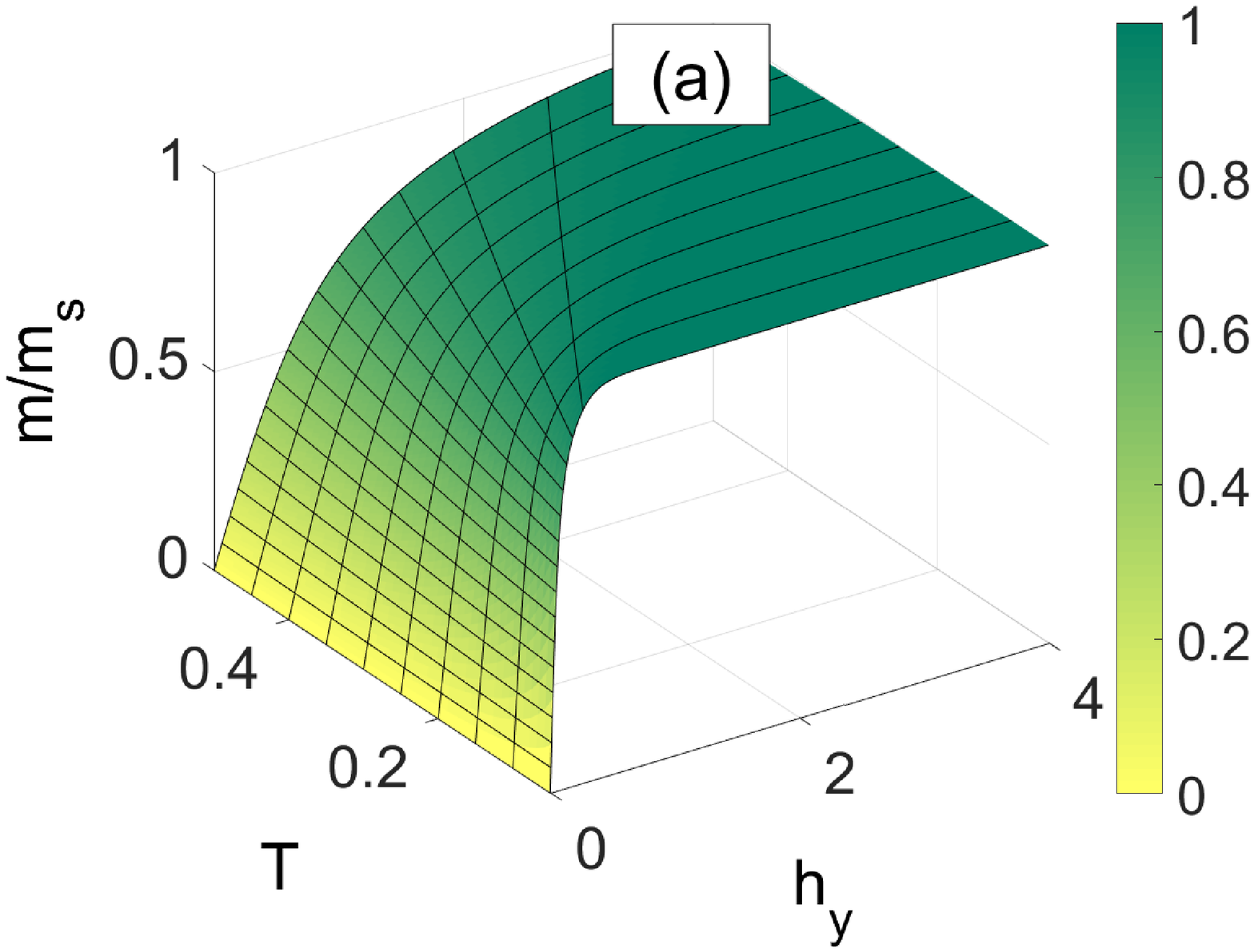}\label{fig:m_FT_SI_J0_hy}}
\subfigure{\includegraphics[scale=0.23,clip]{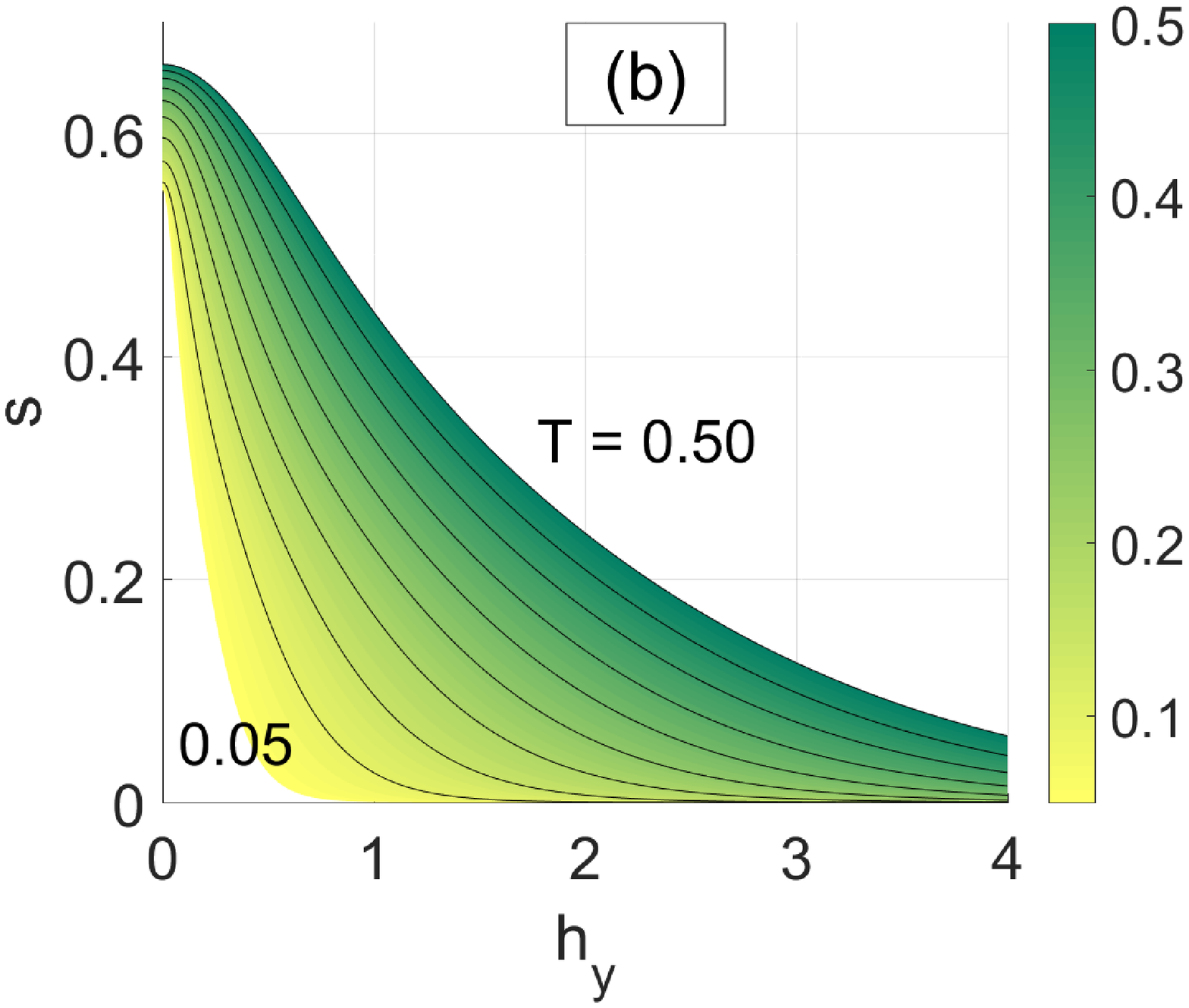}\label{fig:ds_FT_SI_J0_hy}}
\subfigure{\includegraphics[scale=0.23,clip]{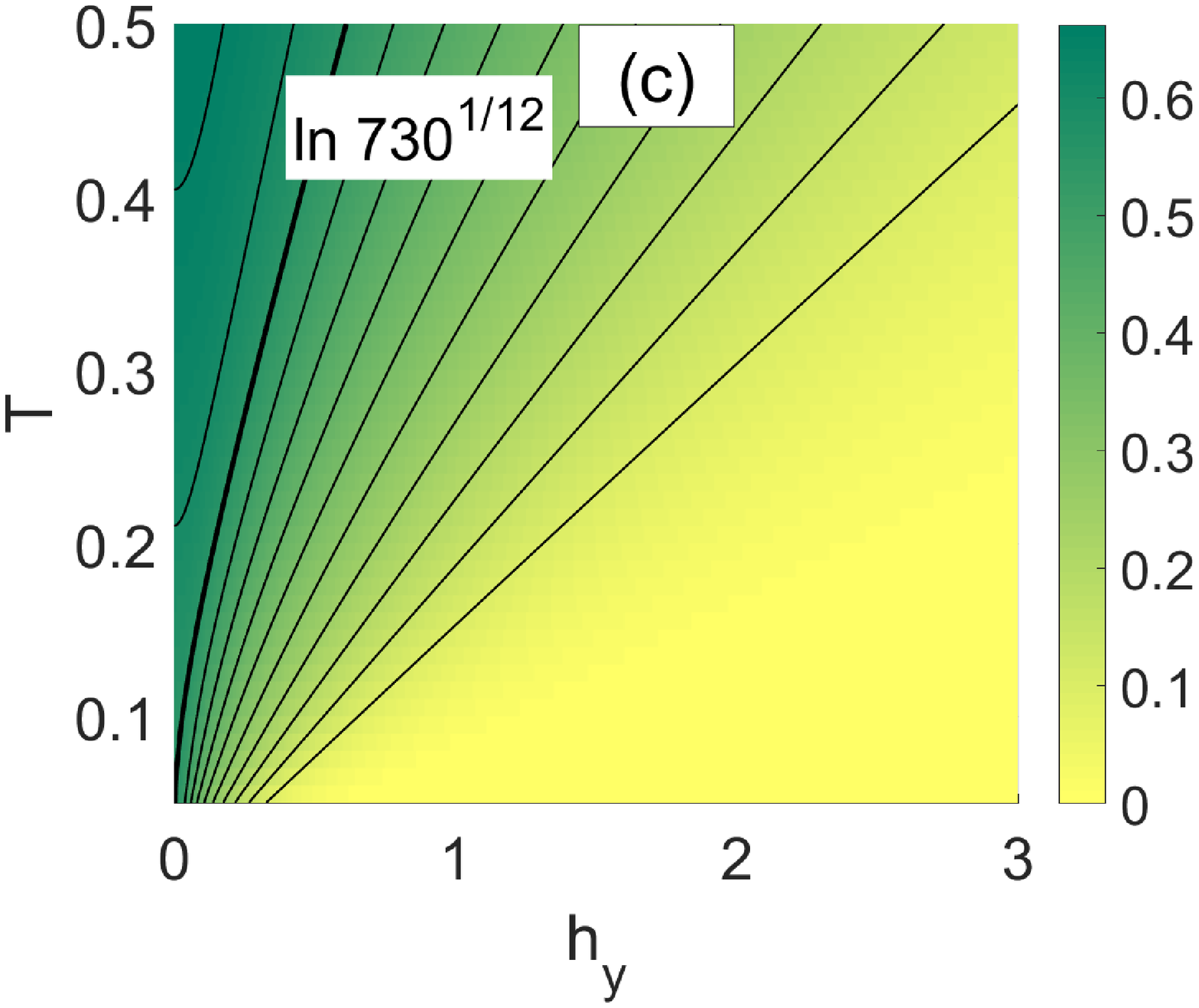}\label{fig:dt_FT_SI_J0_hy}}\\
\subfigure{\includegraphics[scale=0.23,clip]{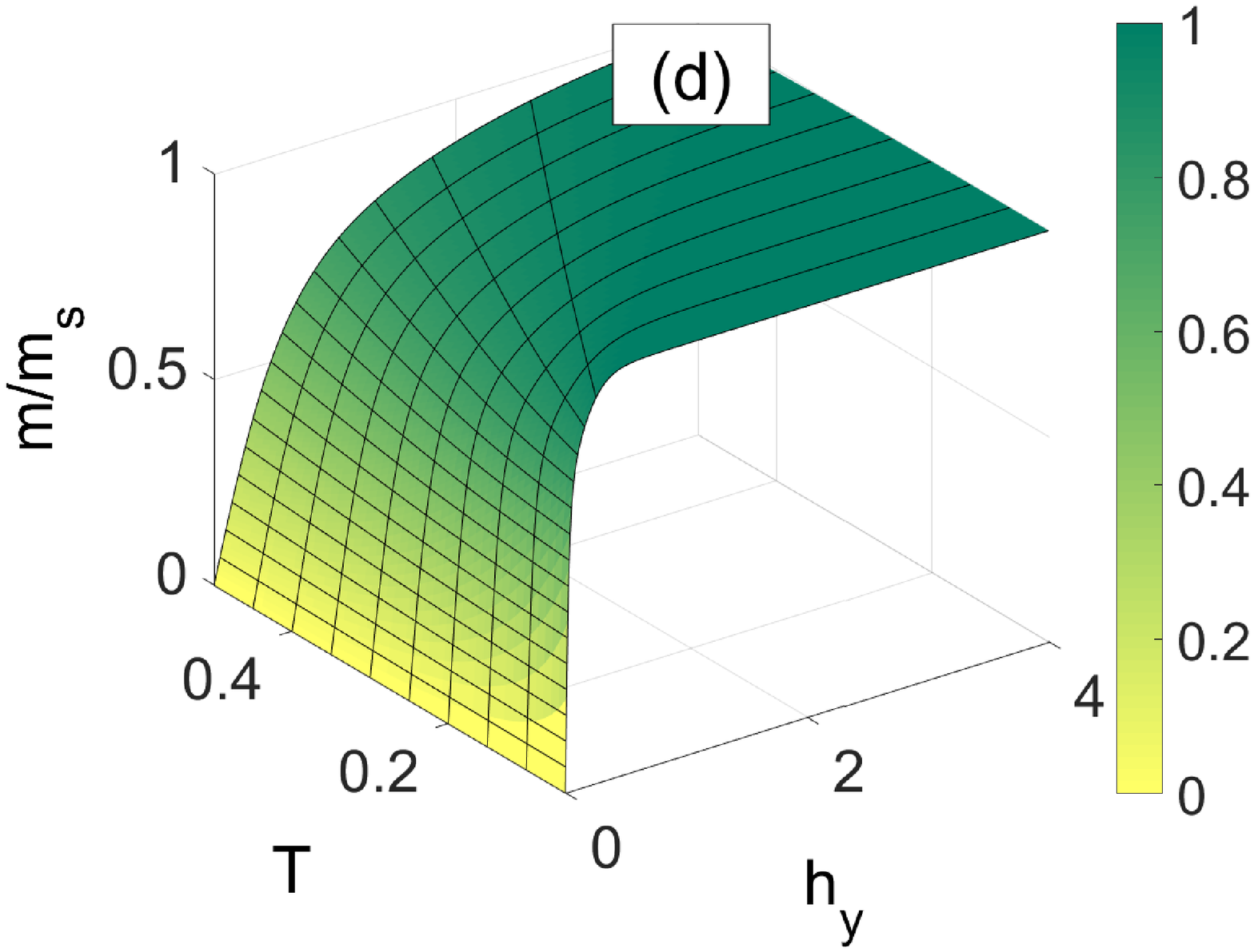}\label{fig:m_FT_SI_J1_hy}}
\subfigure{\includegraphics[scale=0.23,clip]{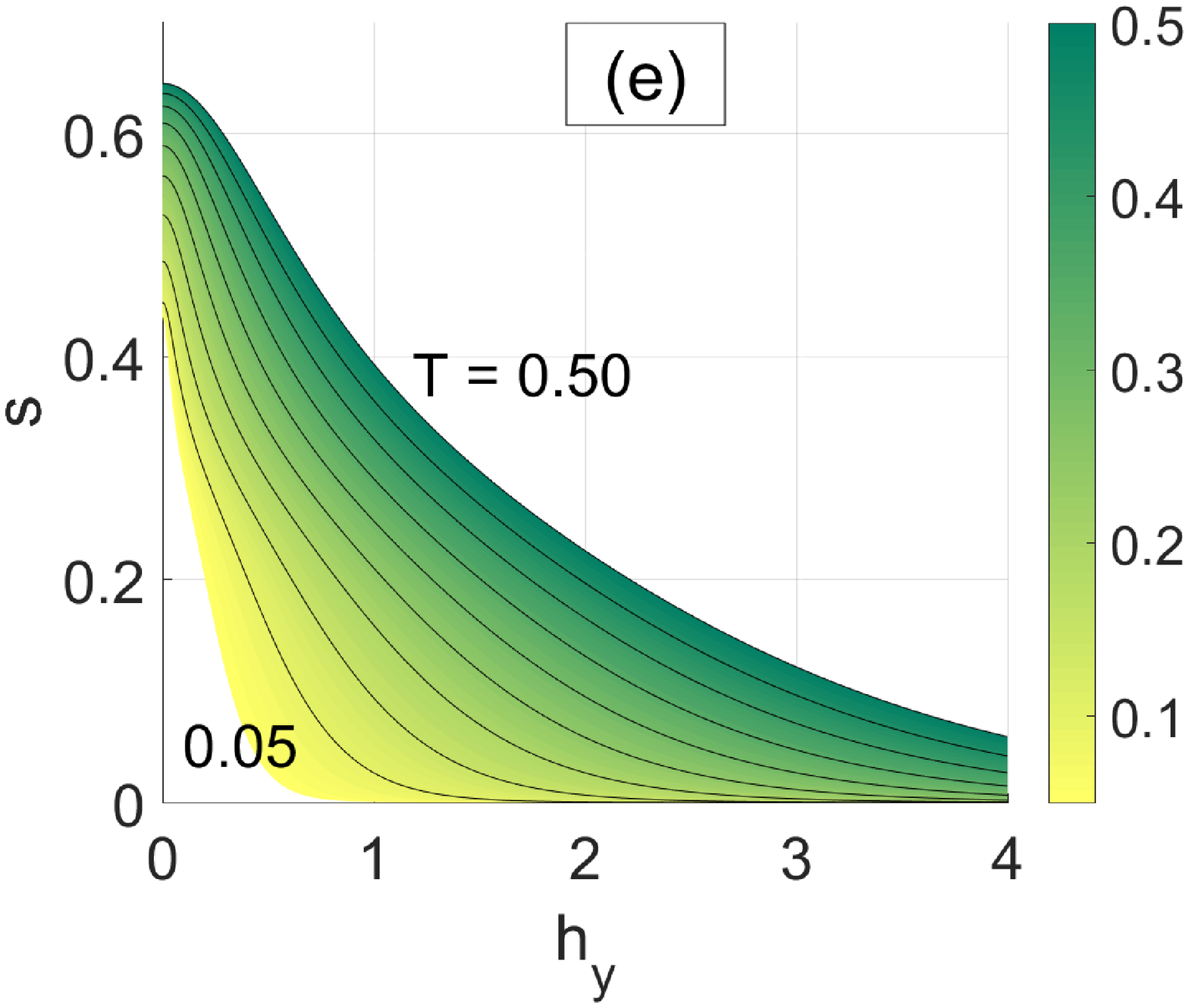}\label{fig:ds_FT_SI_J1_hy}}
\subfigure{\includegraphics[scale=0.23,clip]{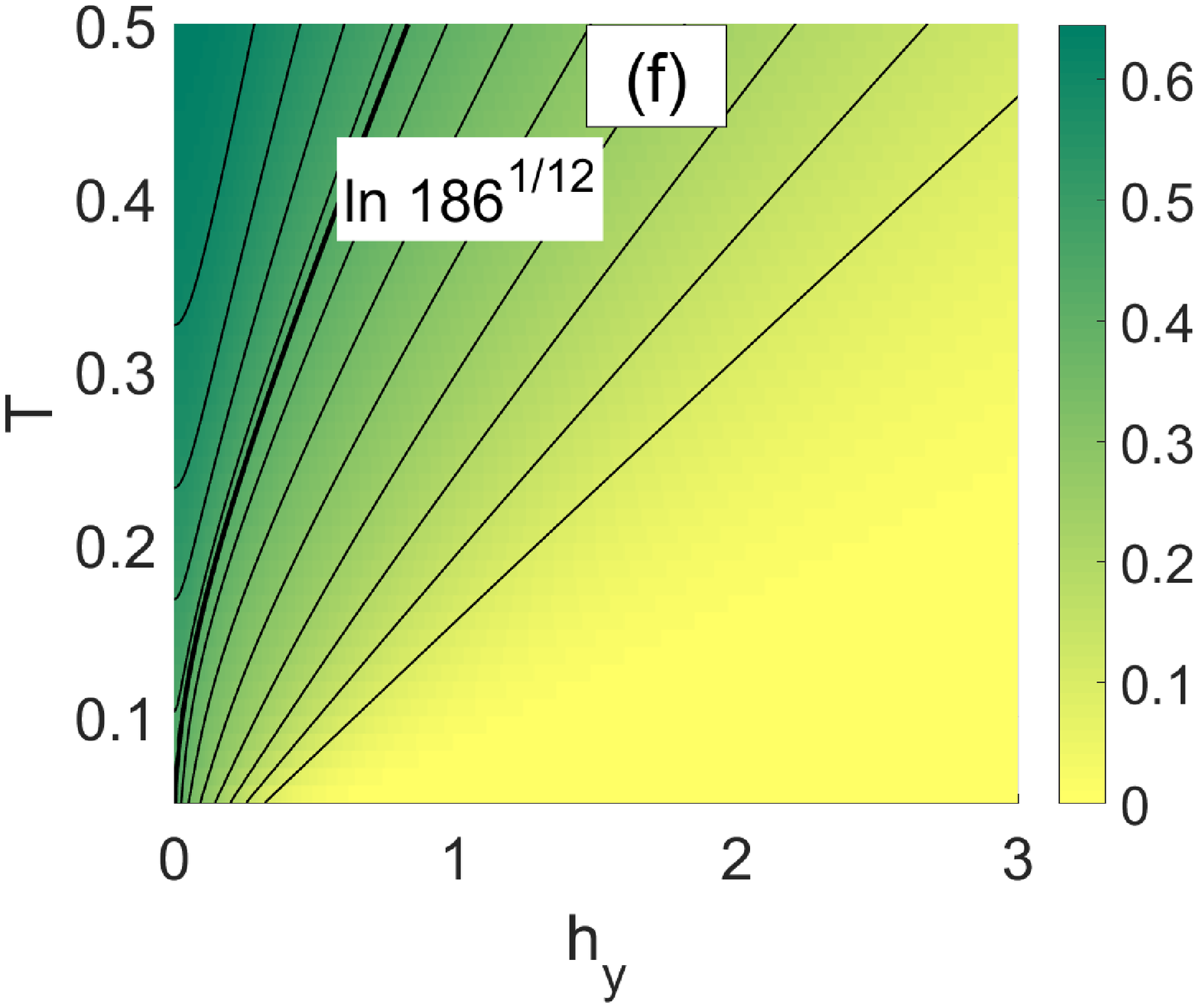}\label{fig:dt_FT_SI_J1_hy}}\\
\subfigure{\includegraphics[scale=0.23,clip]{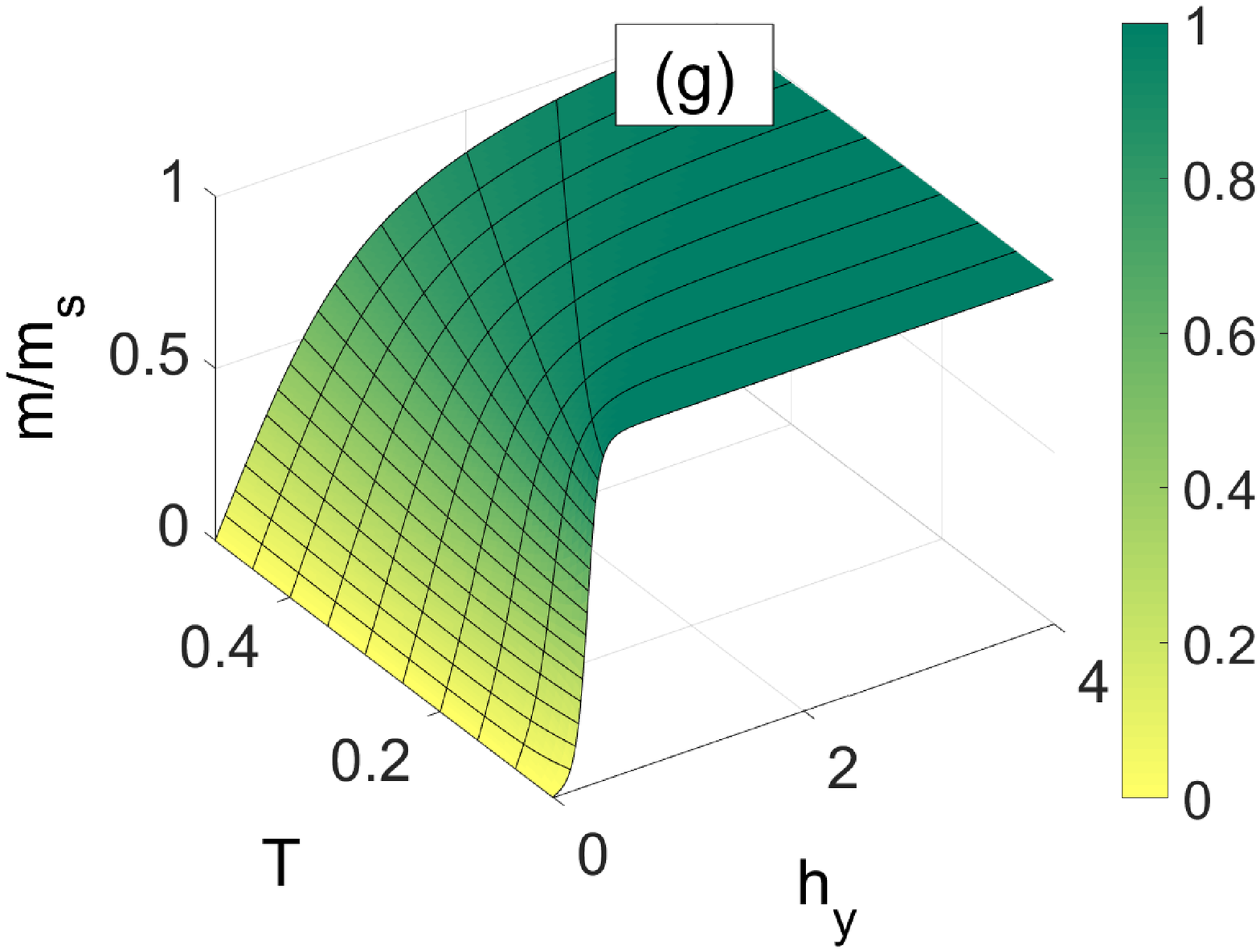}\label{fig:m_FT_SI_J-05_hy}}
\subfigure{\includegraphics[scale=0.23,clip]{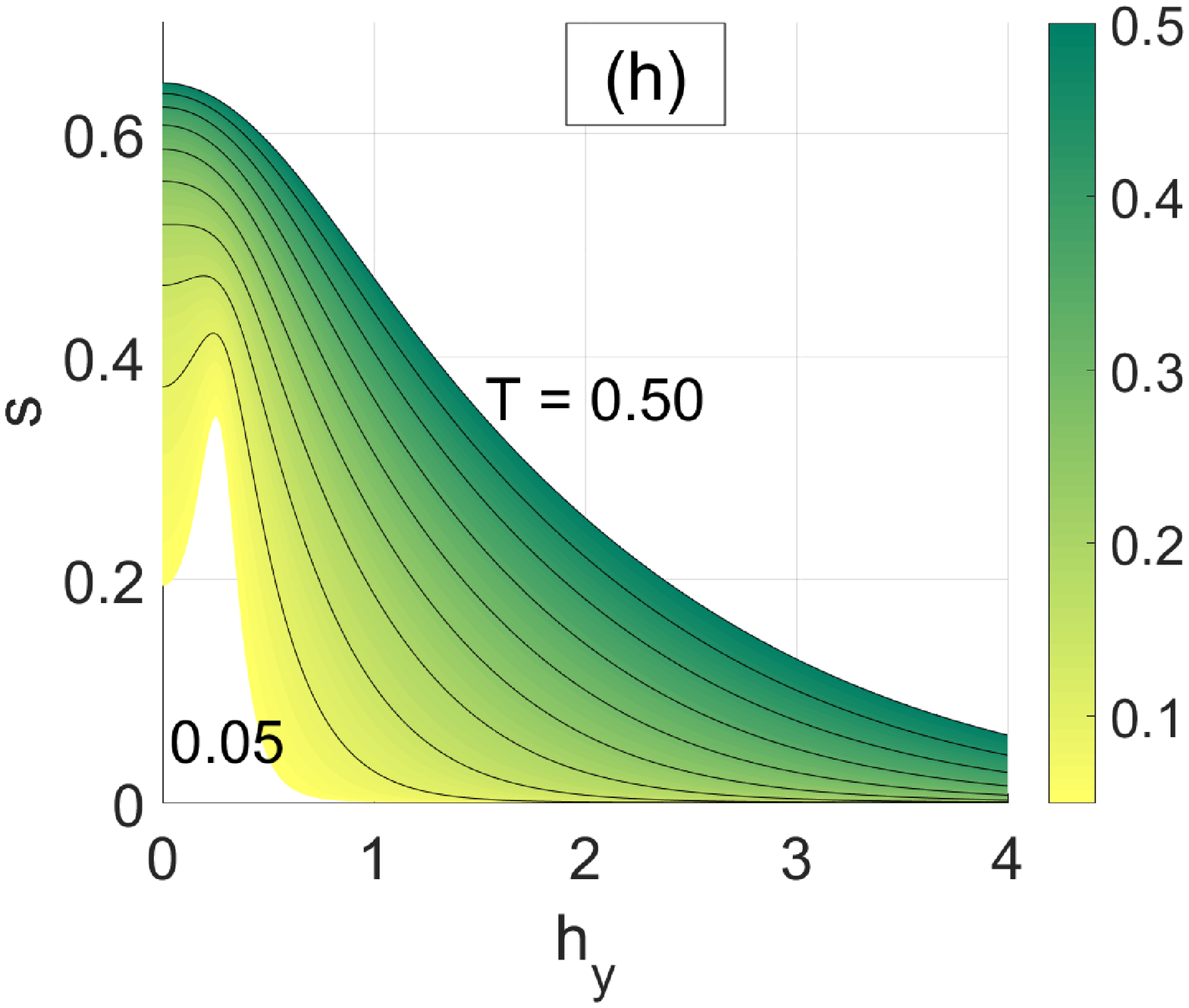}\label{fig:ds_FT_SI_J-05_hy}}
\subfigure{\includegraphics[scale=0.23,clip]{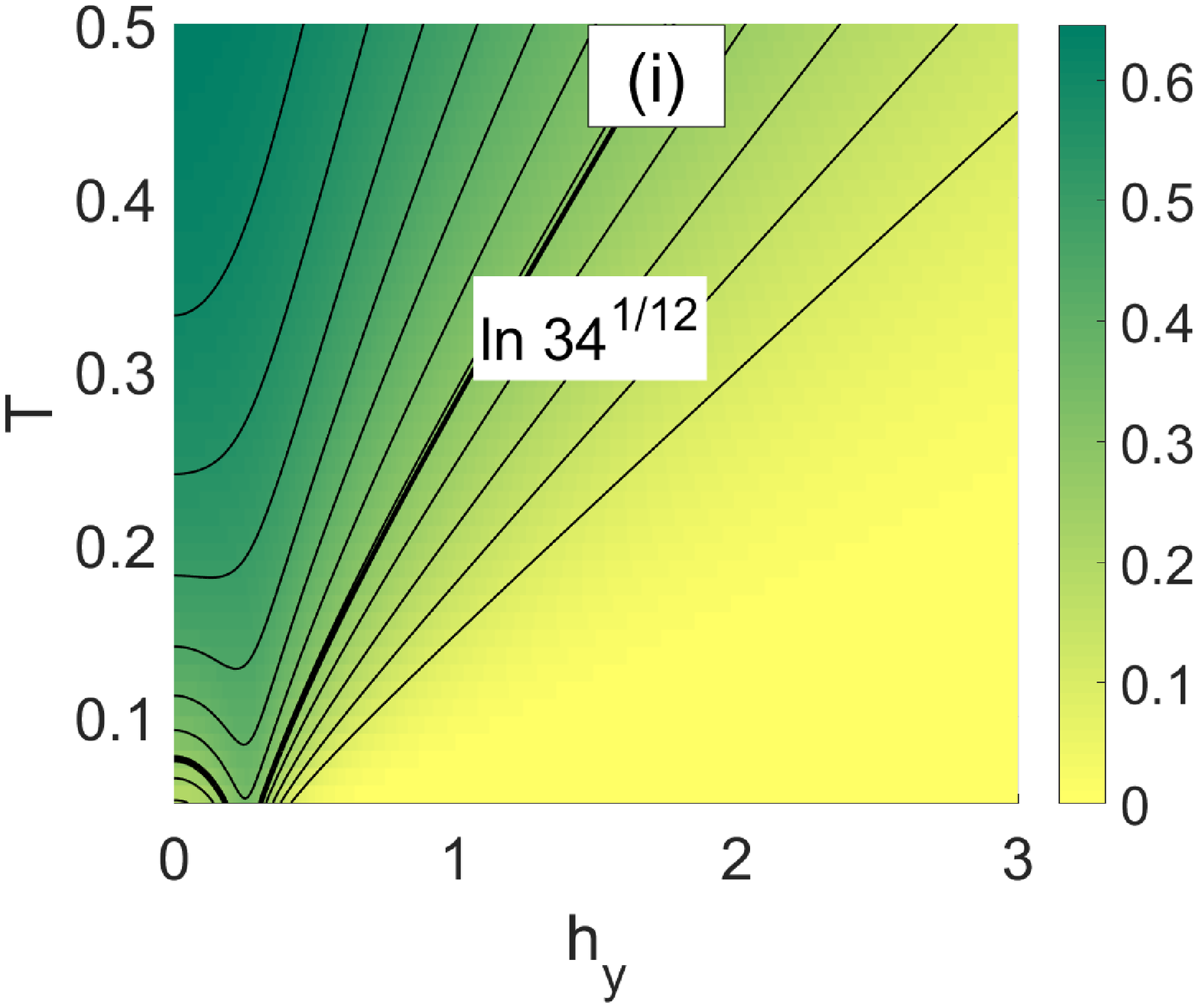}\label{fig:dt_FT_SI_J-05_hy}}
\caption{Spin ice: 3D plots of (a) the reduced magnetization $m/m_{s}$ and (b,c) the entropy density $s$ in the $(T,h)$ parameter plane. In (b) and (c) the black curves represent, respectively, isothermal entropy changes at different temperatures $T \in [0.05, 0.5]$, with the step $\Delta T=0.05$, and adiabatic temperature changes at various $s$ with a varying magnetic field $h_y$, for $J_2=0$. Panels (d-f) and (g-i) show the same quantities as in (a-c), for $J_2=1$ and $-0.5$, respectively.}\label{fig:FT_SI_y}
\end{figure}

\begin{figure}[t]
\centering
\subfigure{\includegraphics[scale=0.23,clip]{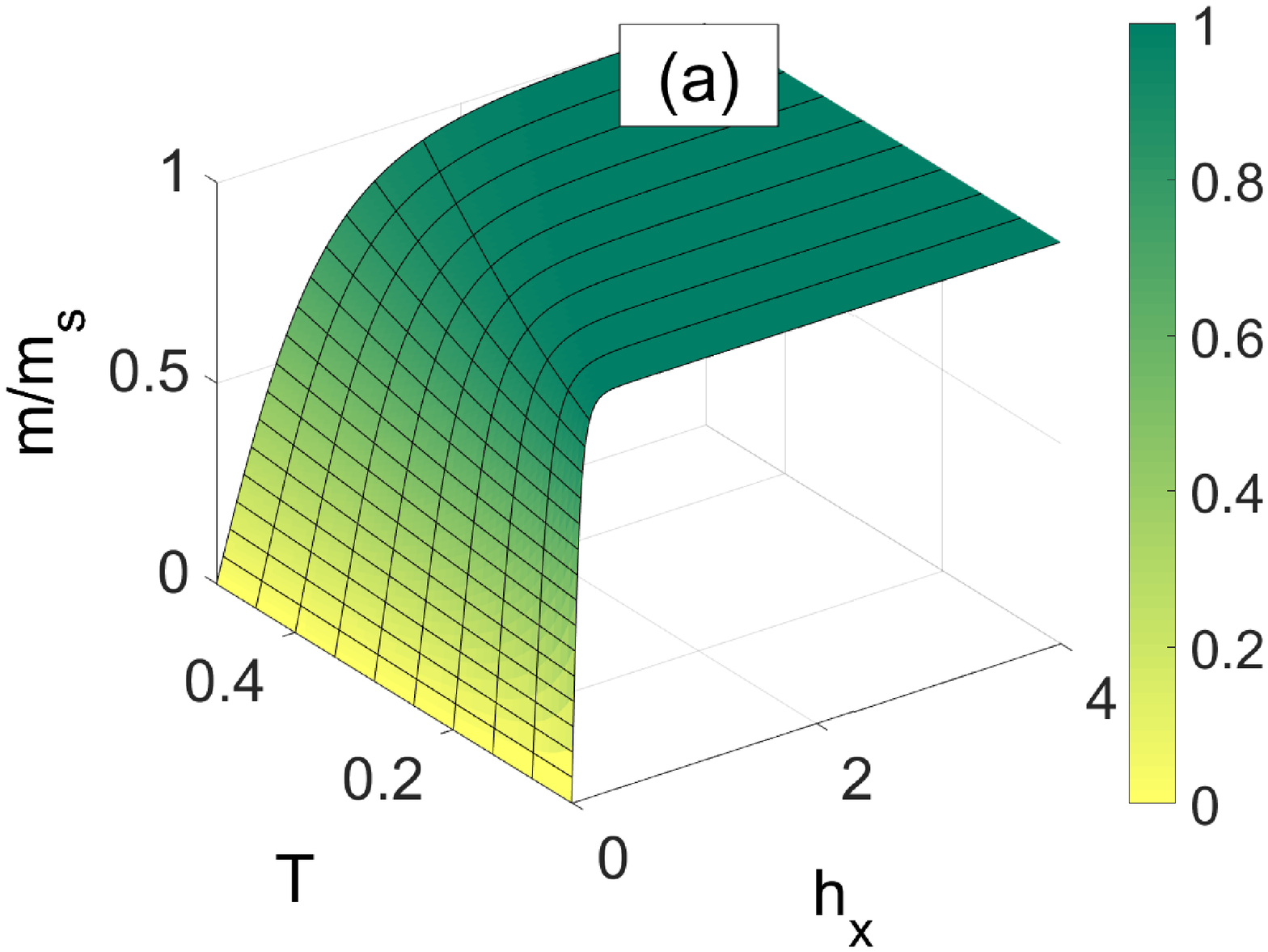}\label{fig:m_FT_SI_J0_hx}}
\subfigure{\includegraphics[scale=0.23,clip]{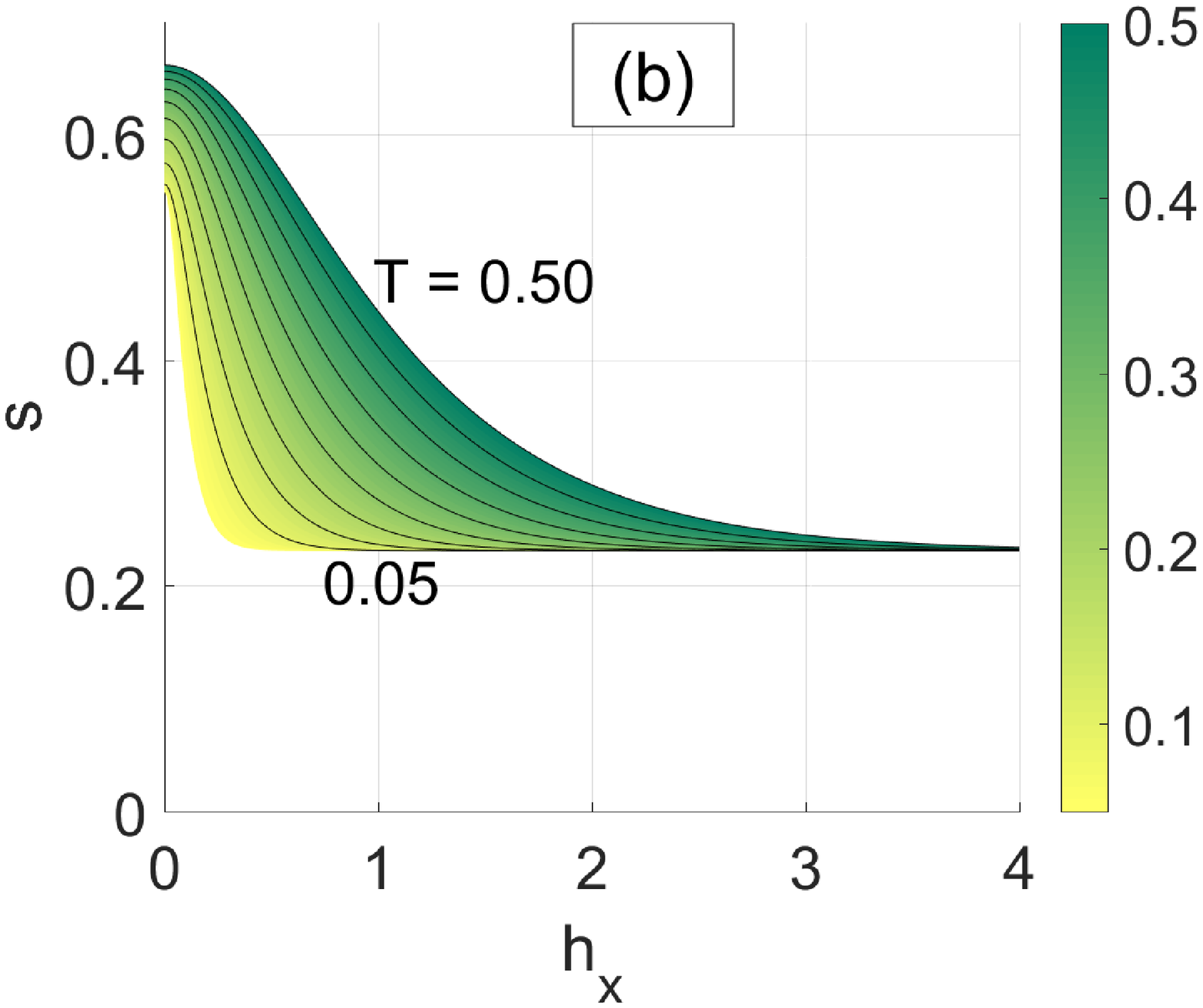}\label{fig:ds_FT_SI_J0_hx}}
\subfigure{\includegraphics[scale=0.23,clip]{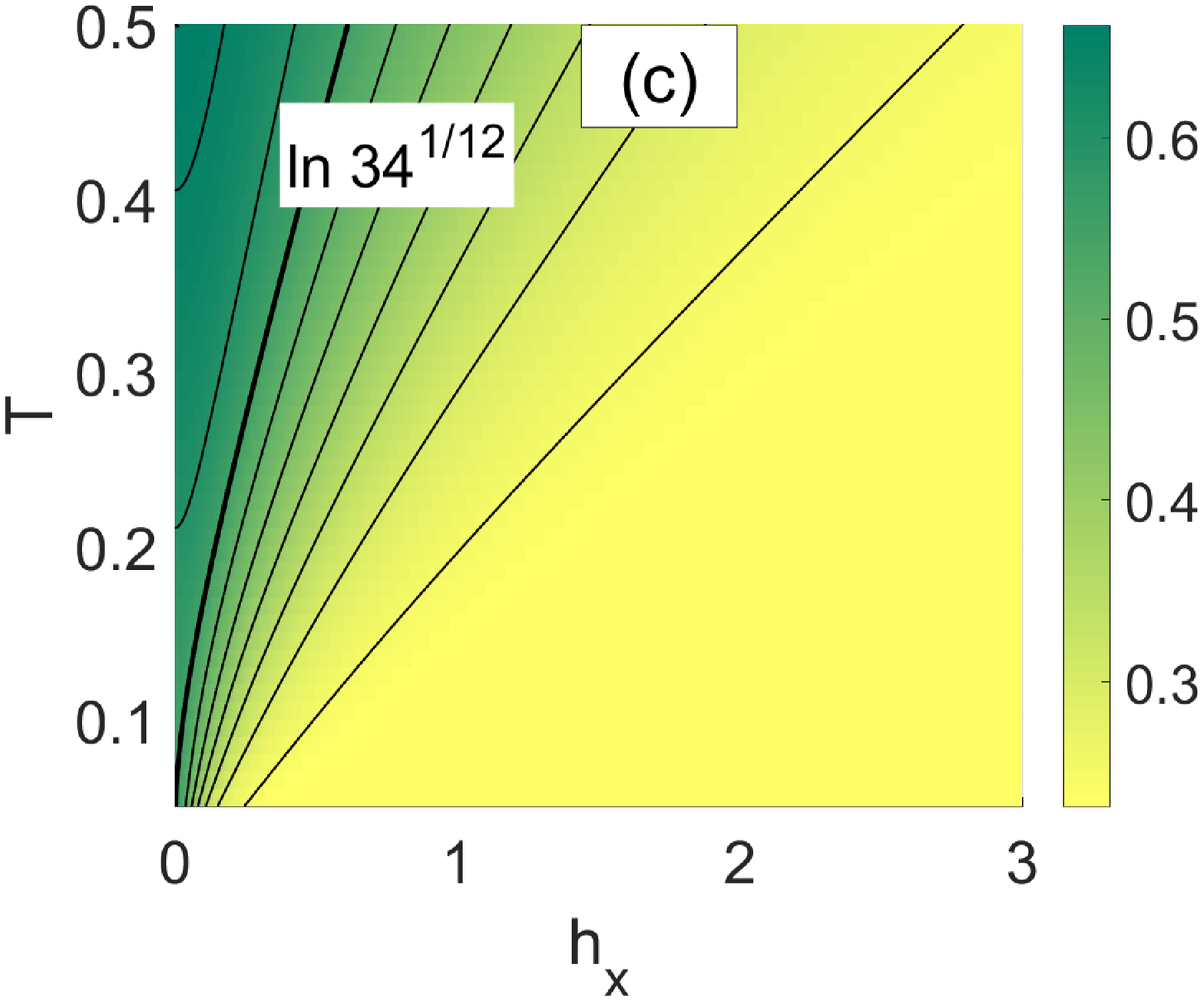}\label{fig:dt_FT_SI_J0_hx}}
\subfigure{\includegraphics[scale=0.23,clip]{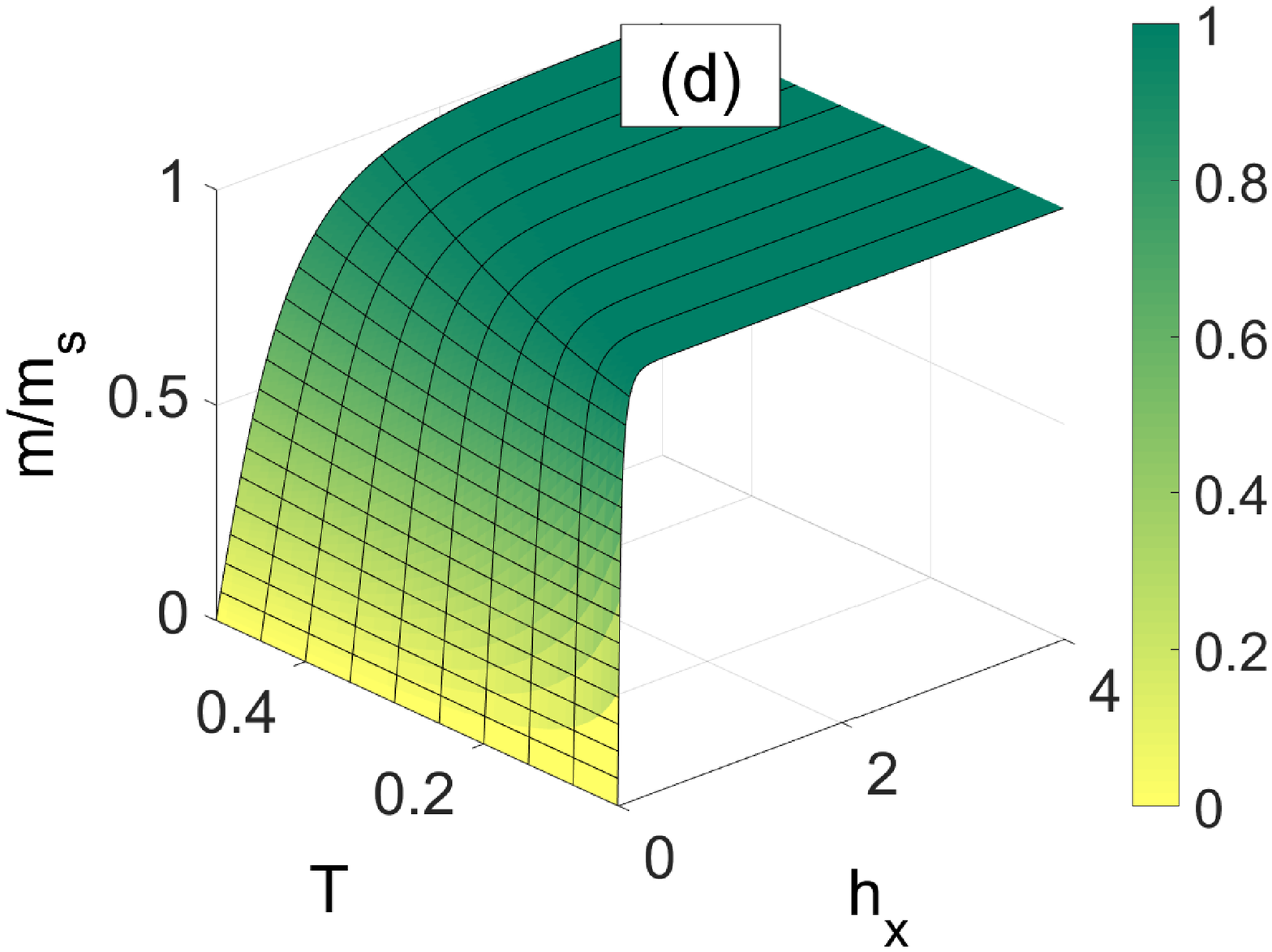}\label{fig:m_FT_SI_J1_hx}}
\subfigure{\includegraphics[scale=0.23,clip]{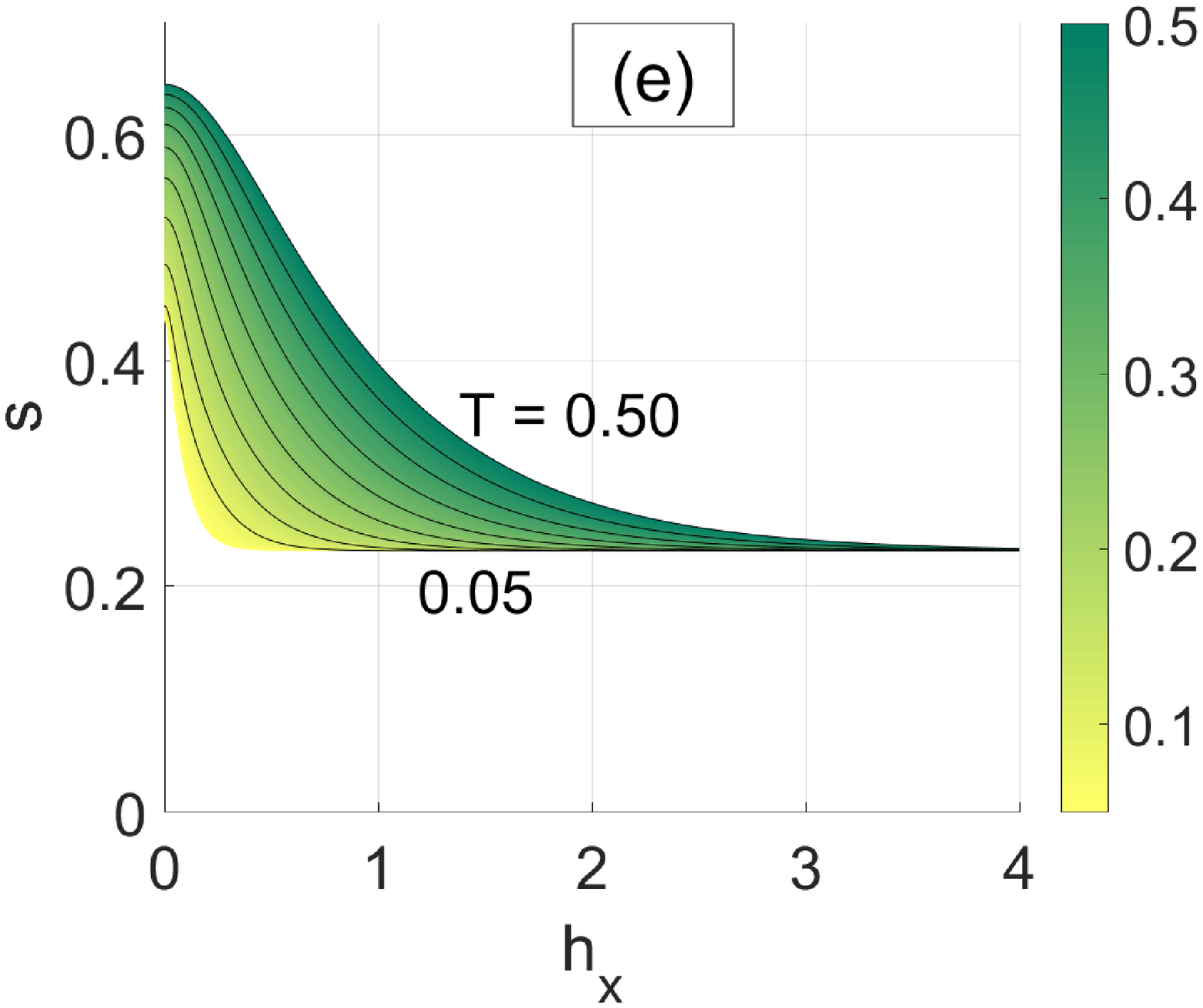}\label{fig:ds_FT_SI_J1_hx}}
\subfigure{\includegraphics[scale=0.23,clip]{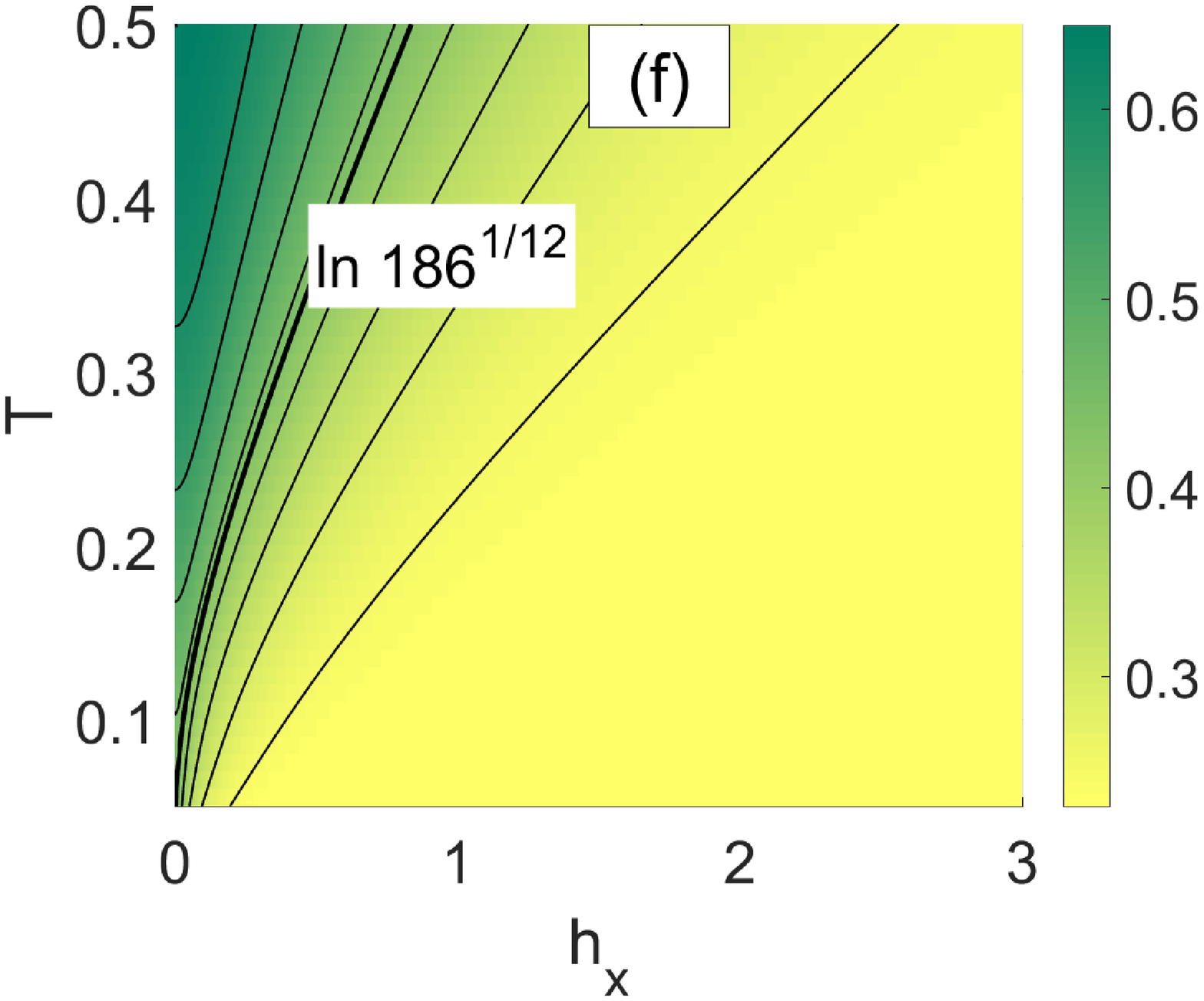}\label{fig:dt_FT_SI_J1_hx}}
\subfigure{\includegraphics[scale=0.23,clip]{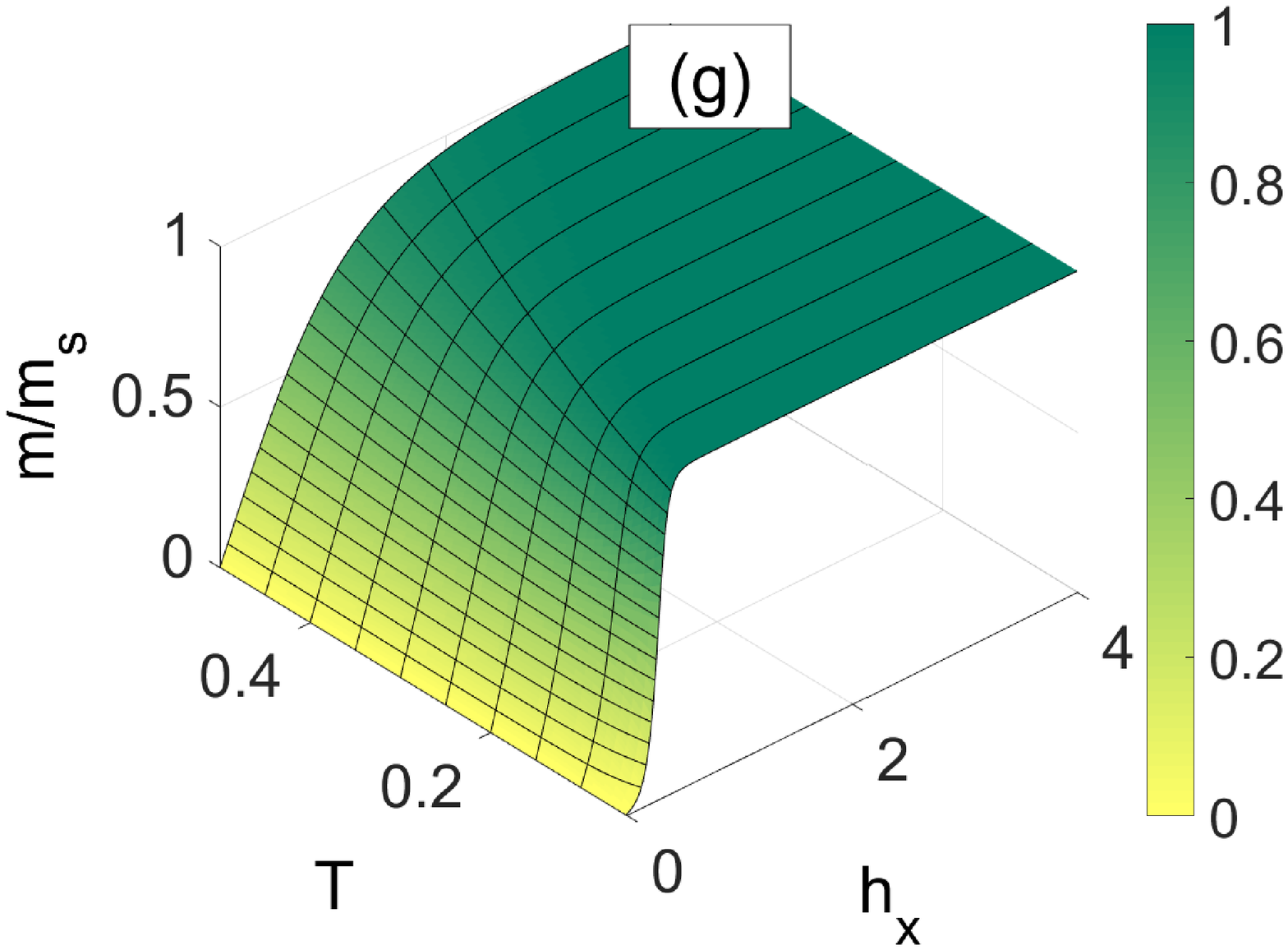}\label{fig:m_FT_SI_J-02_hx}}
\subfigure{\includegraphics[scale=0.23,clip]{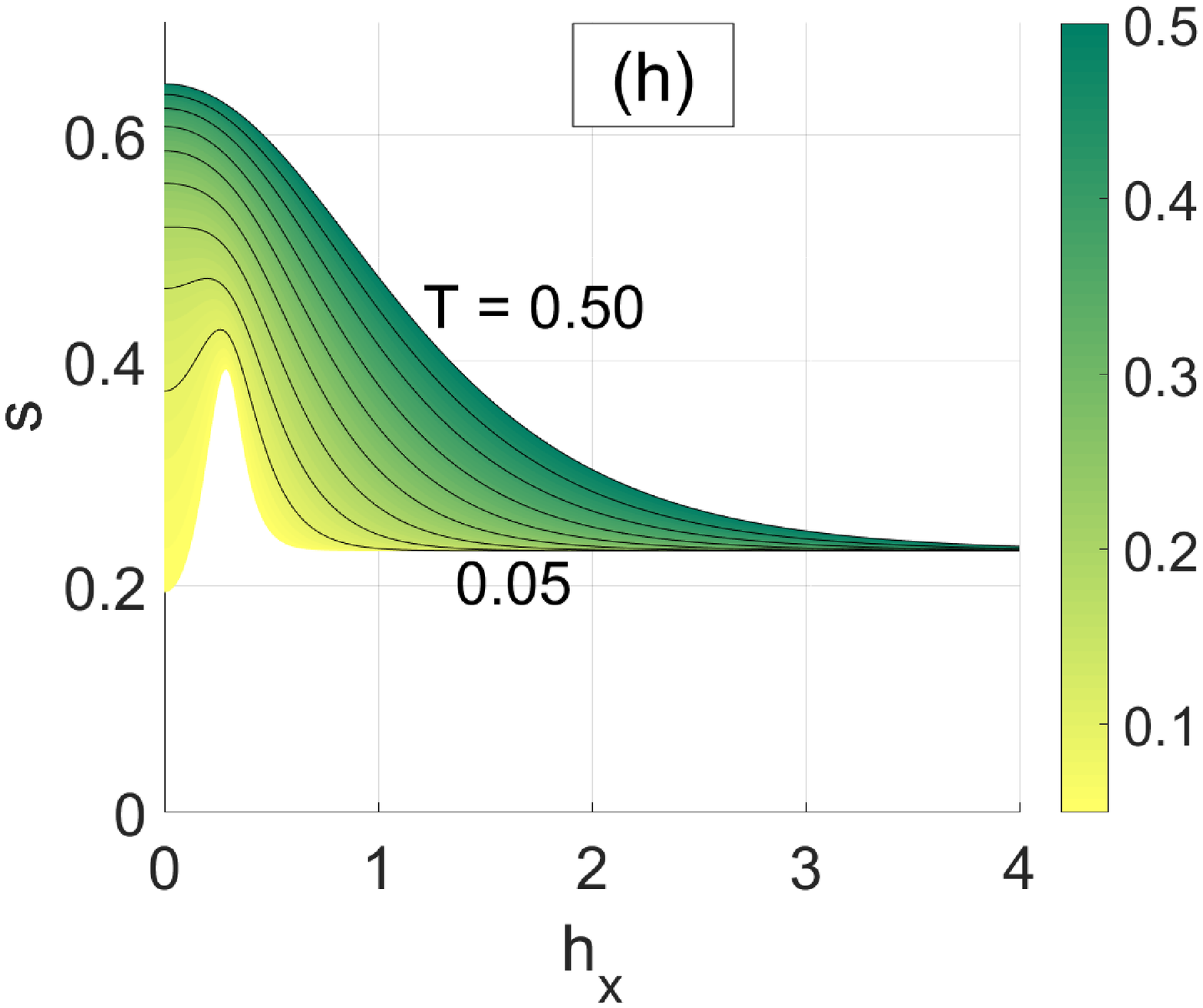}\label{fig:ds_FT_SI_J-02_hx}}
\subfigure{\includegraphics[scale=0.23,clip]{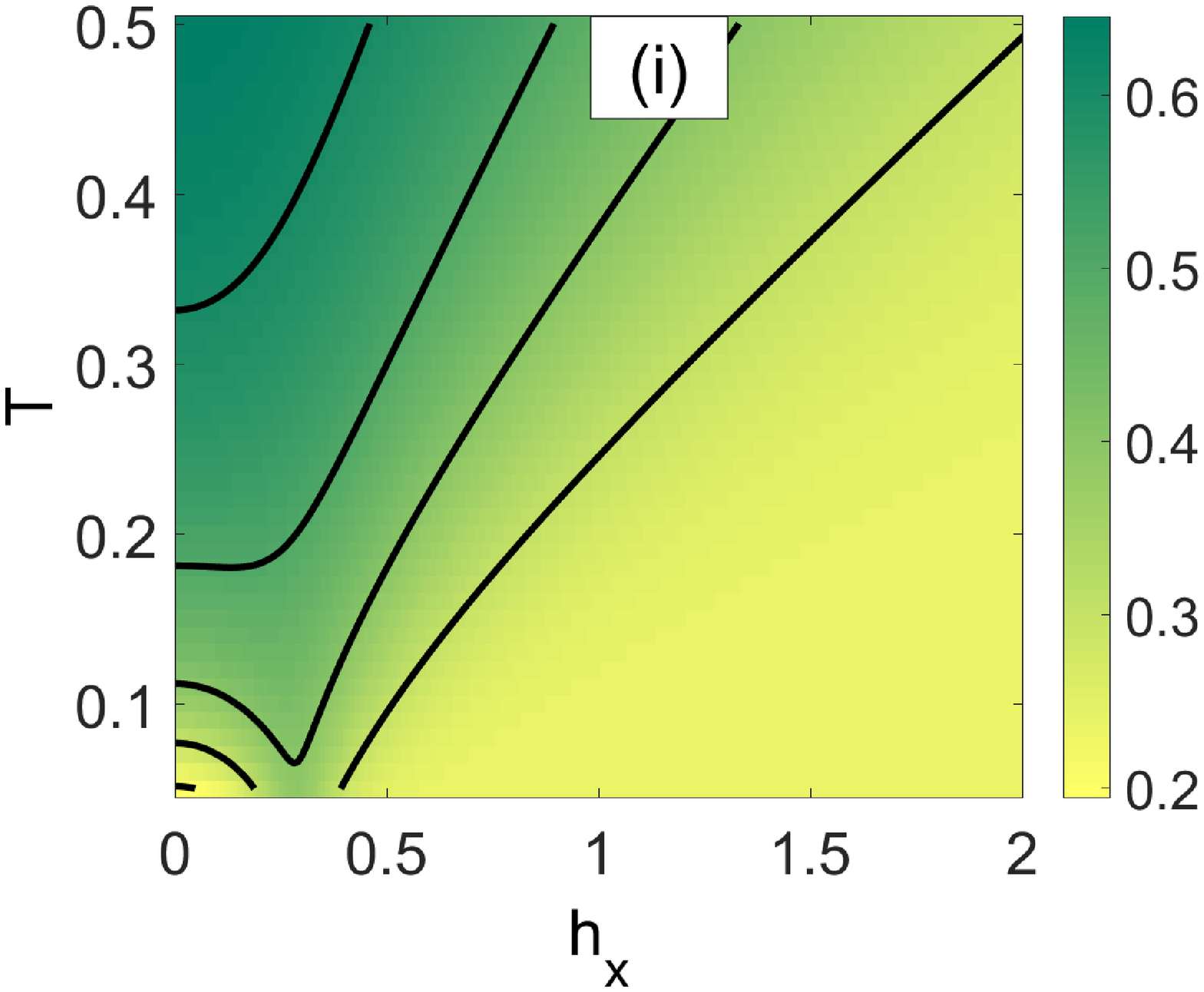}\label{fig:dt_FT_SI_J-02_hx}}
\caption{The same quantities as in Fig.~\ref{fig:FT_SI_y} for the magnetic field $h_x$, applied in the horizontal direction.}\label{fig:FT_SI_x}
\end{figure}

Finite temperature magnetic and magnetocaloric properties, when the field is applied vertically, are presented in Fig.~\ref{fig:FT_SI_y}. In contrast to the IA model, for $J_2>0$ there is no intermediate magnetization plateau and the magnetization increases from zero straight to the saturation value as soon as the magnetic field is turned on. Such a behavior of the magnetization, along with the abrupt entropy decrease from the state with the highest degeneracy in zero field to the nondegenerate state in the saturation phase, translates to an enhanced magnetocaloric effect in the low-temperature adiabatic demagnetization process with the vanishing field. Similar to the IA model, the fastest adiabatic cooling rate at the vanishing field is achieved for the entropies close to the zero-field residual value $s = \frac{\ln 730}{12}$. As shown in the middle and bottom rows of Fig.~\ref{fig:FT_SI_y}, the NNN coupling, both ferromagnetic and antiferromagnetic, diminishes the magnetocaloric effect observed at $J_2 = 0$.\\
\hspace*{5mm} When the magnetic field is applied horizontally, the respective measured quantities show the behavior qualitatively similar to the case when the field is applied in the vertical direction (Fig.~\ref{fig:FT_SI_x}). A peculiar feature is the entropy increase with the increasing field $h_x$ at low-$T$ and a non-vanishing value in the saturation phase.

\section{Summary and conclusion}
\begin{figure}[t]
\centering
\subfigure{\includegraphics[scale=0.4,clip]{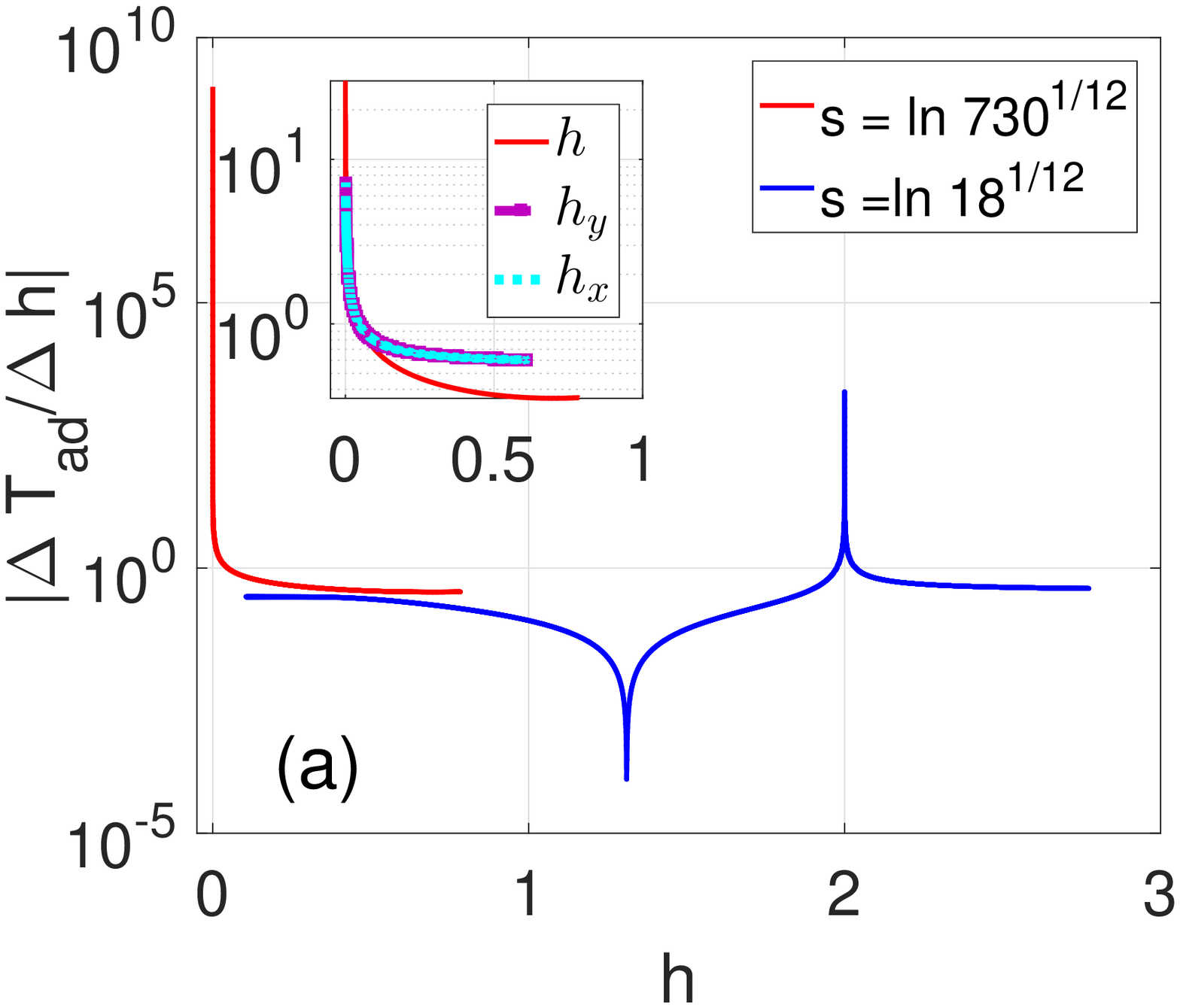}\label{fig:ising_0_zjemneny}}
\subfigure{\includegraphics[scale=0.4,clip]{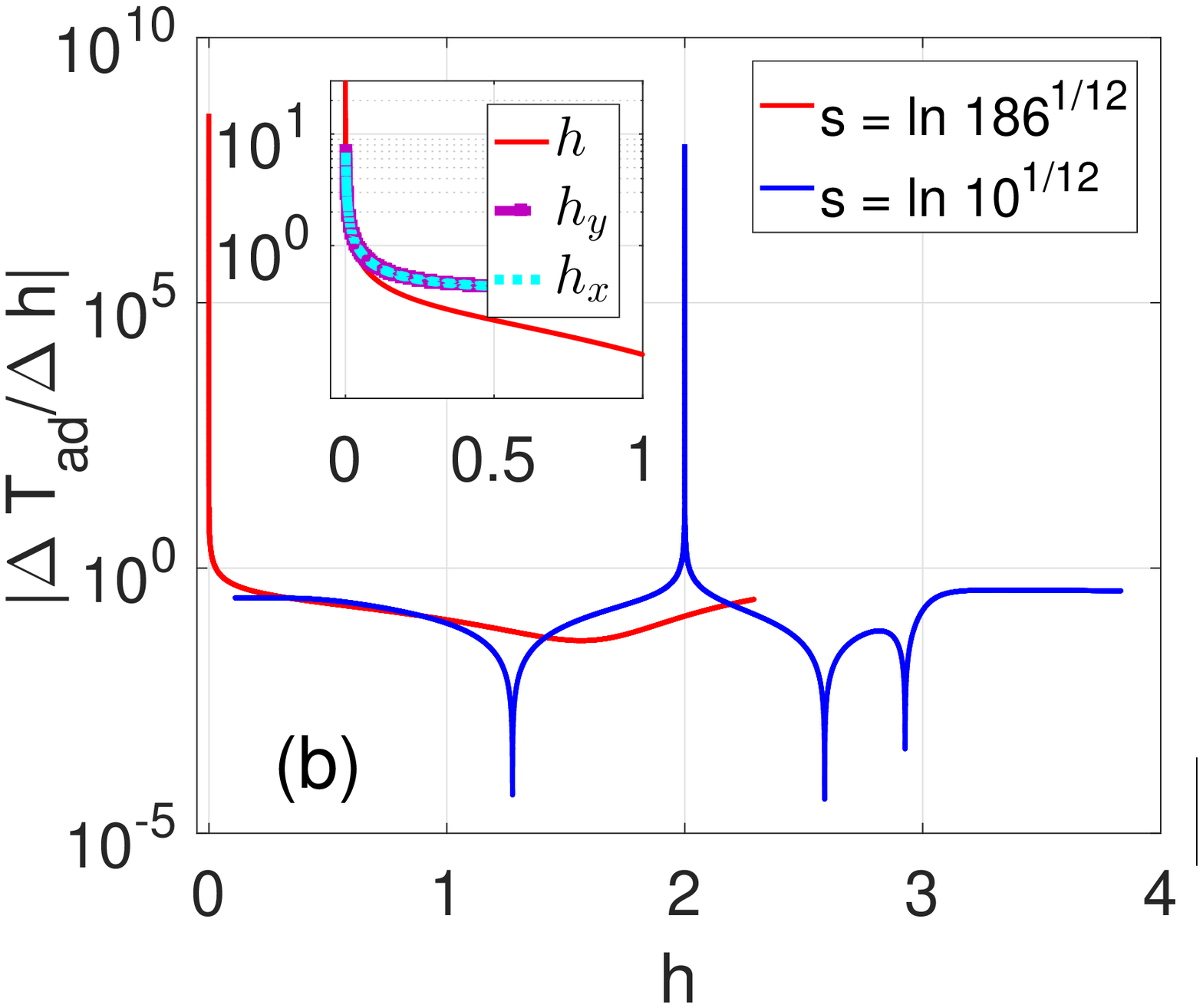}\label{fig:ising_-1_zjemneny}}
\caption{Absolute values of the gradients of the adiabatic temperature changes at the GS entropy values corresponding to the critical fields, for the IA model with (a) $J_2=0$ and (b) $J_2=1$. In (a) the dashed curves correspond to the gradients obtained for the SI model with $J_2=0$ and the fields $h_x$ and $h_y$ tending to zero.}\label{fig:grad}
\end{figure}

We have studied by an exact enumeration magnetic and magnetocaloric properties of two zero-dimensional geometrically frustrated spin systems (nanoclusters): antiferromagnetic Ising (IA) and ferromagnetic spin ice (SI) models, with nearest-neighbor (NN) and the next-nearest-neighbor (NNN) interactions, having a `Star of David' topology. It was found that, depending on the NNN interaction sign and magnitude, the ground state magnetization of the IA model can display one, two or even three intermediate plateaus at fractional values of the saturation magnetization. On the other hand, in the SI model, there is a zero-magnetization (no) plateau for antiferromagnetic (ferromagnetic) NNN couplings, regardless of the applied field direction.\\
\hspace*{5mm} An enhanced magnetocaloric effect has been observed in the vicinity of the respective magnetization jumps. The most prominent one was observed in the adiabatic demagnetization process with the field going to zero in the IA system with $J_2=0$ and $-1$. As demonstrated in Fig.~\ref{fig:grad}, in the adiabatic conditions with the entropy corresponding to the respective zero-field GS degeneracies as the field vanishes the cooling rates go practically to infinity. Similar cooling (heating) rate was also observed for the field approaching the saturation value $h_s=2$ from above (below), in the case of $J_2=-1$ and to less extent also for $J_2=0$. On the other hand, the cooling rates observed in the SI model with $J_2=0$ and the fields applied horizontally and vertically (dashed curves in Fig.~\ref{fig:ising_0_zjemneny}) are comparable with each other but by many orders smaller than in the IA model. Furthermore, except the case of $J_2=J_1$ for the IA model, the presence of NNN interactions in both IA and SI models has an adverse effect on the low-field magnetocaloric properties due to the GS entropy reduction and/or formation of zero-magnetization plateau. \\ 
\hspace*{5mm} Therefore, we can conclude that the `Star of David' nanoclusters with the uniaxial Ising anisotropy and either purely NN or both NN and NNN antiferromagnetic interactions of equal strengths exhibit giant MCE and can be considered for application as efficient low-temperature refrigerants in the adiabatic demagnetization process.

\section*{Acknowledgments}
This work was supported by the grant of the Slovak Research and Development Agency under the contract No. APVV-0132-11 and the Scientific Grant Agency of Ministry of Education of Slovak Republic (Grant No. 1/0331/15). The authors acknowledge the financial support by the ERDF EU (European Union European Regional Development Fund) grant provided under the contract No. ITMS26220120047 (activity 3.2.).

\end{document}